\documentclass[aps,pra,twocolumn]{revtex4-1}

\usepackage{amsmath}
\usepackage{amssymb}
\usepackage{array}
\usepackage{color}
\usepackage{comment}
\usepackage{etoolbox}
\usepackage{graphicx}
\usepackage{multirow}
\usepackage{tikz}
\usepackage{youngtab}

\usepackage[colorlinks,urlcolor=blue,citecolor=blue,linkcolor=blue]{hyperref}

\usetikzlibrary{shapes}

\definecolor{blue}{rgb}{0.29, 0.45, 1.00}
\definecolor{green}{rgb}{0.00, 0.69, 0.31}
\definecolor{red}{rgb}{5.00, 0.10, 0.20}

\newcommand{\n}{\mathbf{n}}

\newcommand{\nn}{\nonumber}
\newcommand{\up}{\uparrow}
\newcommand{\down}{\downarrow}
\newcommand{\bra}[1]{\langle\left.{#1}\right|}
\newcommand{\ket}[1]{\left|{#1}\right.\rangle}

\newcommand{\vect}[1]{{\mathbf#1}}

\newrobustcmd*{\mycircle}[1]{\tikz{\filldraw[draw=#1,fill=#1] (0,0) circle [radius=0.08cm];}}
\newrobustcmd*{\myuptriangle}[1]{\tikz{\node[draw=#1,fill=#1,isosceles triangle,isosceles triangle stretches,shape border rotate=90,minimum width=0.17cm,minimum height=0.17cm,inner sep=0pt] at (0,0) {};}}
\newrobustcmd*{\mydowntriangle}[1]{\tikz{\node[draw=#1,fill=#1,isosceles triangle,isosceles triangle stretches,shape border rotate=270,minimum width=0.17cm,minimum height=0.17cm,inner sep=0pt] at (0,0) {};}}

\begin{document}

\title{SU(N) Fermions in a One-Dimensional Harmonic Trap}

\author{E. K. Laird, Z.-Y. Shi, M. M. Parish and J. Levinsen}

\affiliation{School of Physics and Astronomy, Monash University, Victoria 3800, Australia}

\date{\today}

\begin{abstract}
We conduct a theoretical study of SU($N$) fermions confined by a one-dimensional harmonic potential.  Firstly, we introduce a numerical approach for solving the trapped interacting few-body problem, by which one may obtain accurate energy spectra across the full range of interaction strengths.  In the strong-coupling limit, we map the SU($N$) Hamiltonian to a spin-chain model.  We then show that an existing, extremely accurate ansatz $-$ derived for a Heisenberg SU(2) spin chain $-$ is extendable to these $N$-component systems.  Lastly, we consider balanced SU($N$) Fermi gases that have an equal number of particles in each spin state for $N=2,\,3,\,4$.  In the weak- and strong-coupling regimes, we find that the ground-state energies rapidly converge to their expected values in the thermodynamic limit with increasing atom number.  This suggests that the many-body energetics of $N$-component fermions may be accurately inferred from the corresponding few-body systems of $N$ distinguishable particles.
\end{abstract}

\maketitle

\section{Introduction}
\label{sec:Introduction}

Systems with SU($N$) symmetry are of great importance to many areas of physics.  For example, the SU(2) spin symmetry of electrons is central to the properties of solid-state materials, while the quarks and gluons of quantum chromodynamics transform in representations of the SU(3) color gauge group.  Likewise, the approximately equal interactions between protons and neutrons have inspired the idea that there exists an approximate SU(4) spin-isospin symmetry~\cite{Wigner1937}.  Most recently, interest has been further fueled by the fact that SU($N$) symmetries can be extended to larger $N$ in ultracold atomic gases~\cite{Honerkamp200404, Cazalilla201411}.  Atoms, with their many spin degrees of freedom, can be trapped together in several different internal states using optical manipulation techniques~\cite{Bloch200807}, and a particularly interesting situation occurs for fermionic alkaline-earth atoms~\cite{Fukuhara200701, Boyd200708, Kitagawa200801, Cazalilla200910, Hermele200909, Xu201004, Taie201209, Scazza201408, Zhang201409,Cazalilla201411, Pagano201402,Capponi201604}:\,\,\,In the ground state ($^{1}S_{0}$), these feature zero electronic angular momentum, but non-zero nuclear spin ($F$).  Consequently, there is no hyperfine interaction, and the (nuclear) spin physics is strongly decoupled from the physics of the electron cloud.  When the outer electron densities of two atoms interact, the nuclear spins play no role in the collision, except through Pauli exclusion.  Therefore, the coupling constant $g$ for the interactions is spin-independent, and mathematically, this endows the interaction potential with an SU($N$) symmetry.  Using $^{173}$Yb, different SU($N$)-symmetric states with $N\leq2F+1=6$ have been experimentally realized for fermions confined to a one-dimensional geometry~\cite{Pagano201402} and a three-dimensional lattice \cite{Taie201209};\,\,\,while SU(10)-symmetric fermions have been realized in two-dimensional $^{87}$Sr~\cite{Zhang201409}.

One-dimensional systems, such as the $^{173}$Yb one above, are extremely valuable in the study of many-body physics as they are, in general, more tractable than their higher-dimensional counterparts.  Furthermore, many of them can be solved exactly.  For example, the Bethe Ansatz can be used to solve certain uniform one-dimensional systems of interacting bosons and fermions with an arbitrary number of spin components~\cite{Guan201311}.  Indeed, the experimental realization of one-dimensional SU($N$) Fermi gases~\cite{Pagano201402} has already caused a resurgence of interest in the area from many theoretical groups~\cite{Capponi201604, Decamp201605, Decamp201611, Jiang201603, Beverland201605, Matveeva201606, Pan201702}.  However, exact solutions are not known for interacting systems under harmonic confinement~\cite{Guan201311}.  The importance of this scenario is twofold:\,\,\,firstly, it is the most widely used type of trap in experiments;\,\,\,and secondly, almost any potential can be approximated as a harmonic oscillator at a local minimum.

One way of gaining insight into trapped many-body systems is to probe their behavior in the few-body limit.  Such an approach has already proved to be successful for two-component Fermi gases in a one-dimensional harmonic potential.  Here, a recent experiment~\cite{Wenz201310} investigated the changing interaction energy between a single ``impurity'' fermion (say spin-$\down$) and an increasing number of ``majority'' atoms (say spin-$\up$).  Surprisingly, for weak to moderate interactions, the majority component behaved like a Fermi sea with as few as four $\up$-particles.  On the theoretical side~\cite{Grining201512}, a coupled cluster study on balanced ($N_{\up}=N_{\down}$) systems showed that the ground-state energy rapidly converges to the many-body limit with only a few atoms in each spin state.  Therefore, it is of interest to see whether few SU($N$) fermions can provide similar insight into multi-component gases containing many particles.  Already, for such systems, it has been demonstrated that the many-body contact is well approximated by having just one particle in each component~\cite{Matveeva201606}.

In this paper, we theoretically investigate few-body systems of SU($N$) fermions with short-range interactions in a one-dimensional harmonic trap.  We introduce an efficient scheme for exactly diagonalizing the few-body problem and obtaining the energy spectrum across the full range of interactions.  We furthermore elucidate the underlying group structure of the energy spectra for the three- and four-body problems, and we demonstrate how they can be decomposed into different fermionic and bosonic subsystems.  In the Tonks-Girardeau limit of infinite coupling, we map the SU($N$) problem onto a quantum spin chain, and we show that an approximate analytic expression for the spin-exchange coefficients~\cite{Levinsen201507,Massignan201512} produces extremely accurate results for both the energies and the wave functions of the eigenstates.  Finally, we investigate how the ground-state energy of the SU($N$) system changes as we increase the number of fermions, $n$, in each spin component.  In particular, we show that it rapidly converges with $n$, thus implying that the SU($N$) ground-state properties can be approximately derived from those of $N$ distinguishable particles.

The paper is organized as follows:\,\,\,In Sec.~\ref{sec:Model}, we outline our model of one-dimensional trapped fermions with SU($N$) symmetry.  In Secs.~\ref{sec:Three_SU(3)_Fermions} and~\ref{sec:Four-Body_Problem}, we work through the SU($N$) three- and four-body problems, respectively.  By re-casting the Hamiltonian in a carefully chosen basis, we are able to eliminate two harmonic oscillator indices from the calculation, which simplifies the diagonalization of the resulting matrix-eigenvalue problem.  In Sec.~\ref{sec:Strong-Coupling_Limit}, we focus on the strong-coupling limit, and show that the aforementioned ansatz~\cite{Levinsen201507,Massignan201512} yields energies and wave functions that are essentially indistinguishable from numerically exact results for $N=2,\,3,\,4$.  In Sec.~\ref{sec:Rapid_Convergence_of_the_Ground-State_Energies}, we consider the evolution of the ground-state energy for SU($N=2,\,3,\,4$) Fermi gases with an increasing number of particles of each spin.  Concluding remarks are made in Sec.~\ref{sec:Conclusions_and_Outlook}.

\section{Model}
\label{sec:Model}

The Hamiltonian for trapped, one-dimensional SU($N$) Fermi gases, as realized in cold-atom experiments, is given by
\begin{align} \label{eq:INTRO-01}
\hat{H}
& = \hat{H}_{0} + \hat{H}_{\mathrm{int}} \nn \\
& = \sum_{i\,=\,1}^{\mathcal{N}}\left[-\frac{\hbar^{2}}{2m}\frac{\partial^{2}}{\partial z_{i}^{2}}+\frac{m\omega^{2}}{2}z_{i}^{2}\right]+g\sum_{i\,<\,j}\delta(z_{ij})\,,
\end{align}
where $m$ is the atom mass, $\omega$ is the harmonic oscillator frequency, $z_{ij}\equiv z_i-z_j$ is the relative distance between atoms at positions $z_i$ and $z_j$, and $\mathcal{N}$ is the total number of particles.  The coupling constant $g$ sets the strength of the short-range contact interactions in $\hat{H}_{\mathrm{int}}$, and this can be tuned either by changing the transverse confinement~\cite{Olshanii199808} or, in the SU(2) case, by using a Feshbach resonance.  Note that Pauli exclusion ensures identical fermions do not interact.  In the following, we work in units where $\hbar=m=\omega=1$, such that the harmonic oscillator length $a_{ho}=\sqrt{\hbar/m\omega}=1$.

The single-particle part of the Hamiltonian can be viewed as an $\mathcal{N}$-dimensional harmonic oscillator:
\begin{align} \label{eq:2}
\hat{H}_0 = - \frac{1}{2} \partial_{\vect{z}}^T\partial_{\vect{z}} + \frac{1}{2}  \vect{z}^T \vect{z}\,,
\end{align}
where we have the row vectors $\partial_{\vect{z}}^T = \{\partial_{z_1},\,\partial_{z_2},\,\ldots,\,\partial_{z_\mathcal{N}} \}$ and $\vect{z}^T = \{z_1,\,z_2,\,\ldots,\,z_\mathcal{N}\}$.  This non-interacting Hamiltonian has eigenstates $\ket{\vect{n}} \equiv \ket{n_1,\,n_2,\,\ldots,\,n_{\mathcal{N}}}$, and corresponding energy eigenvalues $\sum_i \left(n_i +\frac{1}2\right)$, where $n_{i}$ is the harmonic oscillator quantum number of the $i^{th}$ atom.

\begin{figure*}[t]
\begin{center}
\includegraphics[scale=0.3]{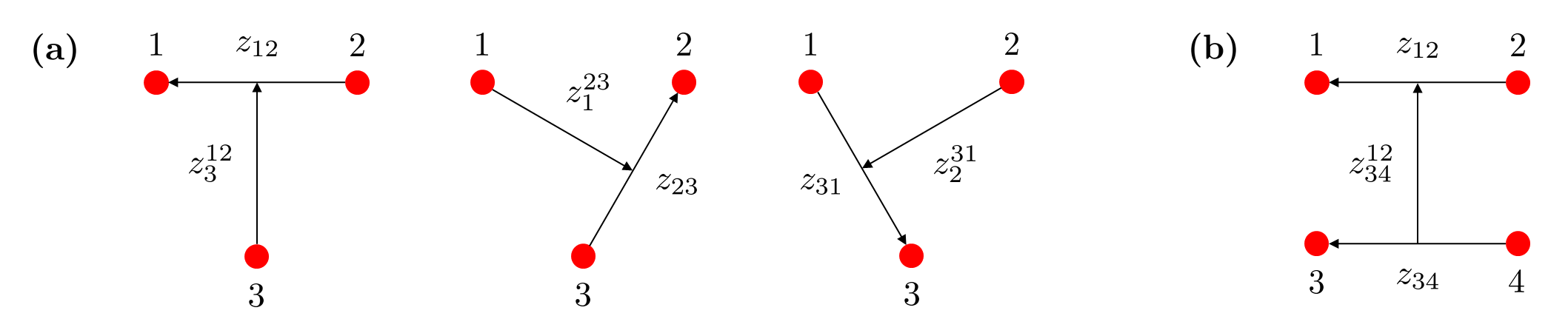}
\caption{Sketch of the coordinate transformations:\,\,\,\textbf{(a)}\,\,\,In the three-body problem, we consider the motion of a pair of atoms with the relative coordinate $z_{ij}$, and the relative motion between the pair's centre of mass and the third atom with the coordinate $z_{k}^{ij}$.  When we take the delta-function interaction boundary condition on the wave function, we are left with a single degree of freedom $-$ the relative atom-pair motion.\,\,\,\textbf{(b)}\,\,\,To deal with the introduction of a fourth particle, we look at the relative motion of two pairs through $z_{ij}$ and $z_{kl}$, and additionally, the relative motion between their two centres of mass through $z_{kl}^{ij}$ (see also Ref.~\cite{Bradly2014}).  After taking the boundary condition, we are left with two motional degrees of freedom.}
\label{fig:sketch}
\end{center}
\end{figure*}

A general interacting $\mathcal{N}$-body state can be written as a superposition of the non-interacting eigenstates:
\begin{align} \nn
|\psi^{(\mathcal{N})}\rangle=\sum_{\{n_i\}}\phi_{n_1n_2\ldots n_\mathcal{N}}
|n_1,\,n_2,\,\ldots,\,n_{\mathcal{N}}
\rangle\equiv\sum_\n \phi_\n|\mathbf{n}
\rangle \,,
\end{align}
with the amplitude $\phi_{n_1n_2\ldots n_\mathcal{N}} \equiv \phi_\n=\langle{\n}|\psi^{(\mathcal{N})}\rangle$.  By considering successive pairs of coordinates, as shown in Fig.~\ref{fig:sketch}, it is possible to perform a coordinate transformation $\vect{z}' = U^{ij}\vect{z}$, such that a given interaction $g\delta(z_{ij})$ only depends on one coordinate, while $\hat{H}_0$ remains a set of decoupled one-dimensional harmonic oscillators.  In other words, we may write
\begin{align}
\hat{H}_0 =  - \frac{1}{2} \partial_{\vect{z}'}^T \mathcal{D}\partial_{\vect{z}'} + \frac{1}{2}  \left(\vect{z'}\right)^T \mathcal{D}^{-1} \vect{z}'\,,
\end{align}
where $\mathcal{D} \equiv U^{ij} \left(U^{ij}\right)^T$ is a diagonal matrix, and the transformed set of coordinates $\vect{z}'$ contains $z_{ij}$.  This allows us to substantially simplify the exact diagonalization of the interacting $\mathcal{N}$-particle problem.

In the following sections, we first use a system of three distinguishable fermions to demonstrate our method in detail.  Then we present the results for the four-body problem, which can be solved in an analogous manner.

\section{Three-Body Problem}
\label{sec:Three_SU(3)_Fermions}

We begin our few-body study by considering three distinguishable SU($3$) fermions in a one-dimensional harmonic trap.  By transforming to the centre-of-mass and relative coordinates of the particles, we obtain the full energy spectrum, and show that it can be understood in terms of the spectra for one- and two-component Bose and Fermi gases.  Unlike the uniform case, exact solutions for harmonically confined one-dimensional systems are not generally known.  Accordingly, the three-body problem has been treated numerically using various approaches:\,\,\,see, for example, Refs.~\cite{Gharashi201307, Astrakharchik201308, Loft201605} for two-component fermions, and Refs.~\cite{Harshman201211, DAmico201402, Garcia-March2014} where all three atoms interact.

As outlined in the previous section, a general three-body state may be written as
\begin{align} \label{eq:INTRO-02}
|\psi^{(3)}\rangle=\sum_{n_1,\,n_2,\,n_3%,\,n_{2},\,n_{3}
}\phi_{n_1n_2n_3}|n_{1},n_{2},n_{3}
\rangle\equiv\sum_\n%,\,n_{2},\,n_{3}
\phi_\n|\mathbf{n}%n_{1},n_{2},n_{3}
\rangle\,,
\end{align}
with the amplitude $\phi_{n_{1}n_{2}n_{3}}\equiv \phi_\n=\langle{\n}|\psi^{(3)}\rangle$.  Taking the time-independent Schr\"{o}dinger equation $(E-\hat{H})|\psi^{(3)}\rangle=0$, and projecting it onto the non-interacting eigenstate $|n_{1},n_{2},n_{3}\rangle\equiv\ket{\n}$, we obtain
\begin{align} \label{eq:TSU(3)F-01}
\left[E-(n_{1}+n_{2}+n_{3})\right]\phi_\n&\nn \\
&\hspace{-30mm}=g\sum_{\n'}\bra{\n}\left[\delta(z_{12})+\delta(z_{23})+\delta(z_{31})\right]\ket{\n'}\phi_{\n'}\,,
\end{align}
where the total energy $E$ is measured with respect to the zero-point energy of the harmonic oscillators.

Reminiscent of the two-body interactions that are occurring in the gas, we then rotate our coordinates so that instead of looking at the motion of individual particles, we consider the motion of different types of ``pairs'' (see Fig.~\ref{fig:sketch}).  Therefore, in addition to the relative coordinate $z_{ij}$, we consider the relative motion of the centre of mass of a pair with an atom, $z_{k}^{ij}\equiv(z_{i}+z_{j})/2-z_{k}$, as well as the centre-of-mass motion of all three atoms,  $z_{\textit{cm}}\equiv(z_{1}+z_{2}+z_{3})/3$.  Associated with each of these coordinates ``$z$'' is a harmonic oscillator quantum number ``$n$'', and since the energy is unchanged by the coordinate transformation, we require:\,\,\,$n_{1}+n_{2}+n_{3}=n_{\textit{cm}}+n_{k}^{ij}+n_{ij}$.  As we shall see, the advantage of this transformation is that instead of having an expression, Eq.~\eqref{eq:TSU(3)F-01}, involving three quantum numbers $\{n_{1},\,n_{2},\,n_{3}\}$, each summed from $0$ to $\infty$, we eventually obtain a matrix equation in terms of a single index (namely, $n_{k}^{ij}$).

To proceed, we insert a complete set of transformed states $|n_{cm},n_{k}^{ij},n_{ij}\rangle$ on either side of the operators $\delta(z_{ij})$ in Eq.~\eqref{eq:TSU(3)F-01}.  Since the interaction only changes the quantum number for ``atom-atom'' motion $n_{ij}$, we have $\langle n_{cm},n_{k}^{ij},n_{ij}|\delta(z_{ij})|n_{cm},n_{k}^{ij},n_{ij}'\rangle=\varphi_{n_{ij}}\varphi_{n_{ij}'}$, where
\begin{align} \label{eq:TSU(3)F-02}
\def\arraystretch{1.4}
\varphi_{n_{ij}}=
\left\{
\begin{array}{ll}
(-1)^{n_{ij}/2}\sqrt{\sqrt{\dfrac{1}{2\pi}}\dfrac{(n_{ij}-1)!!}{n_{ij}!!}}\,,\,\,\, & n_{ij}\,\,\mathrm{even} \\
0\,,\,\,\, & n_{ij}\,\,\mathrm{odd}
\end{array}
\right.
\end{align}
is the relative harmonic oscillator eigenfunction for particles $i$ and $j$ at zero separation.  Taking the boundary condition $z_{ij}=0$, and also setting $n_{cm} = 0$ (since the interaction energy is independent of the centre-of-mass motion), the wave function coefficients $\phi$ can be replaced by new coefficients $\eta$:
\begin{align} \label{eq:TSU(3)F-03}
\eta_{ij}\equiv\eta_{(ij)}^{n_{k}^{ij}}=g\sum_{n_{ij},\,\n}\varphi_{n_{ij}}\langle0,n_{k}^{ij},n_{ij}|\n\rangle\phi_\n\,,
\end{align}
where $\{i,\,j,\,k\}=\{1,\,2,\,3\}$ and cyclic permutations.

Using Eqs.~\eqref{eq:TSU(3)F-02} and~\eqref{eq:TSU(3)F-03}, Eq.~\eqref{eq:TSU(3)F-01} becomes
\begin{align} \label{eq:TSU(3)F-04}
\left[E-(n_{1}+n_{2}+n_{3})\right]\phi_\n
& = \sum_{n_{3}^{12},\,n_{12}}\varphi_{n_{12}}\langle\n|0,n_{3}^{12},n_{12}\rangle\eta_{12}\nn \\
& +\sum_{n_{1}^{23},\,n_{23}}\varphi_{n_{23}}\langle\n|0,n_{1}^{23},n_{23}\rangle\eta_{23}\nn \\
& +\sum_{n_{2}^{31},\,n_{31}}\varphi_{n_{31}}\langle\n|0,n_{2}^{31},n_{31}\rangle\eta_{31}\,.
\end{align}
Continuing, we divide by $E-(n_{1}+n_{2}+n_{3})$, and then act with the operator
\begin{align} \label{eq:TSU(3)F-05}
g\sum_{n_{ij}',\,\n}\varphi_{n_{ij}'}\langle0,n_{k}^{ij\prime},n_{ij}'|\n\rangle(\,\cdot\,)
\end{align}
on the left, three separate times, where $\{i,\,j,\,k\}$ take the same values as on the right-hand side of Eq.~\eqref{eq:TSU(3)F-04}.  This yields three equations, one for each $\eta_{ij}$.  Making use of the identity,
\begin{align} \label{eq:TSU(3)F-06}
\sum_{\n}|\n\rangle\frac{1}{E-(n_{1}+n_{2}+n_{3})}\langle\n|=\frac{1}{E-\hat{H}_{0}}\,,
\end{align}
where $\hat H_0$ is the non-interacting Hamiltonian, we eventually obtain a matrix equation:
\begin{align} \label{eq:TSU(3)F-07}
\frac{1}{g}
\left(\begin{array}{c}
\eta_{12} \\
\eta_{23} \\
\eta_{31} \\
\end{array}\right)
=
\left(\begin{array}{ccc}
A & B & B \\
B & A & B \\
B & B & A \\
\end{array}\right)
\left(\begin{array}{c}
\eta_{12} \\
\eta_{23} \\
\eta_{31} \\
\end{array}\right).
\end{align}
Above, $\{\eta_{ij}\}$ are vectors and $\{A,\,B\}$ are square matrices given by
\begin{align} \label{eq:TSU(3)F-08A}
A_{n,n'}
& = \sum_{l}\frac{\varphi_{l}^{2}}{E-n-l}\delta_{n,n'}\nn \\
& = -\frac{\Gamma[-(E-n)/2]}{2\sqrt{\,2\,}\,\Gamma[1/2-(E-n)/2]}\delta_{n,n'}
\end{align}
and
\begin{align} \label{eq:TSU(3)F-08B}
B_{n,n'}=\sum_{l,\,l'}\frac{\varphi_{l}\varphi_{l'}}{E-n-l}T^{n,l}_{n',l'}\,,
\end{align}
where the matrix indices refer to the ``atom-pair'' quantum number $n_{k}^{ij}$.

\begin{figure*}[t]
\begin{center}
\includegraphics[scale=0.42]{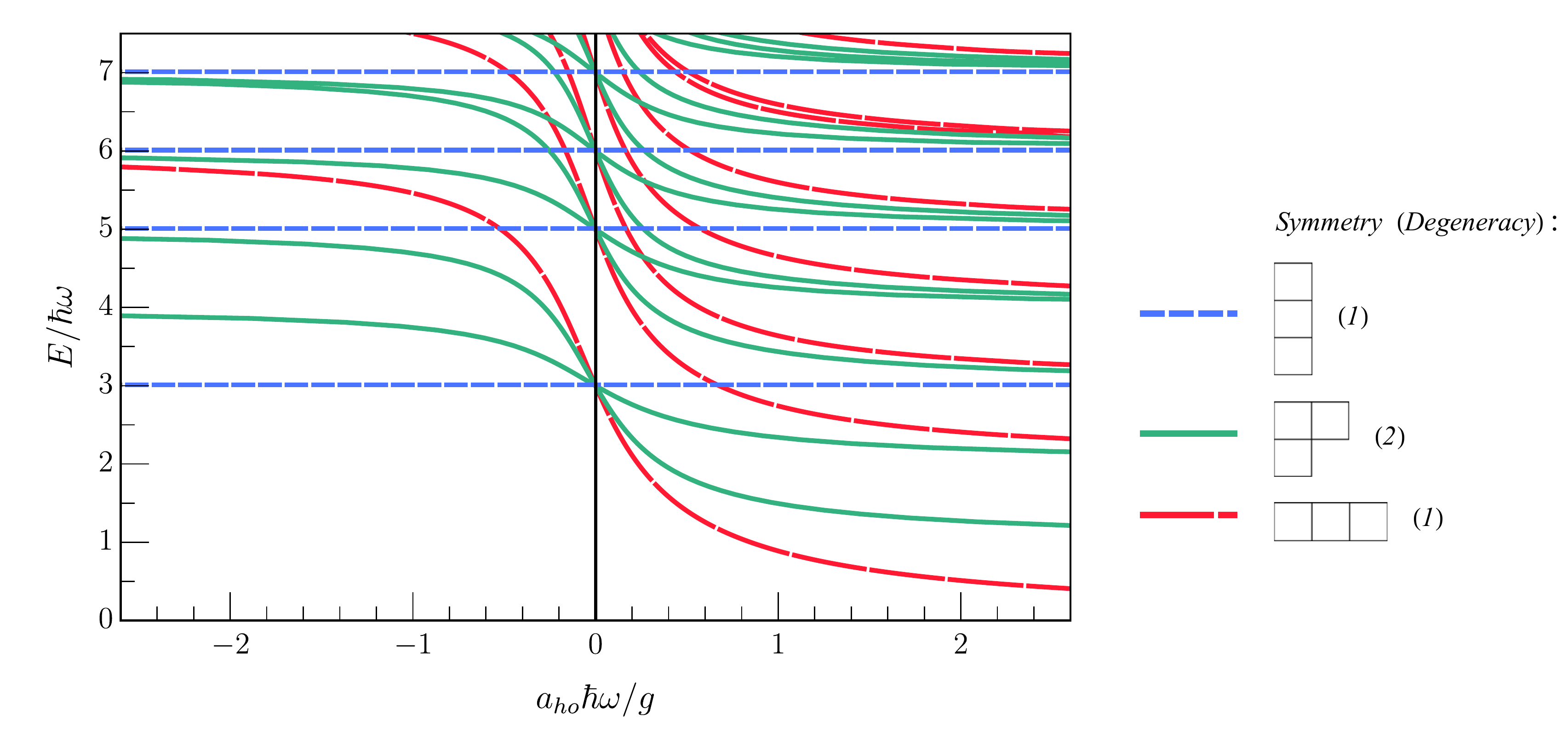}
\caption{The energy spectrum of upper-branch states for three distinguishable SU($3$) fermions interacting in a one-dimensional harmonic trap.  The energies $E$ are found using Eqs.~\eqref{eq:TSU(3)F-07}-\eqref{eq:TSU(3)F-08B}, and are plotted against the inverse interaction strength $1/g$.  This result can be decomposed into
subsystems with fewer components, associated with the irreducible representations of S$_{3}$ (see main text).}
\label{fig:TSU(3)F}
\end{center}
\end{figure*}

Inside $B$, we have the three-body Clebsch-Gordan coefficient $T^{n,l}_{n',l'}$, which is defined via the transformation,
\begin{align}
|0,n_{j}^{ki},n_{ki}\rangle = \sum_{n_{k}^{ij},\,n_{ij}} T^{n_{j}^{ki},n_{ki}}_{n_{k}^{ij},n_{ij}} |0,n_{k}^{ij},n_{ij}\rangle\,,
\end{align}
where $\{i,\,j,\,k\}$ again take the same values as in Eq.~\eqref{eq:TSU(3)F-04}.  Note that the quantum numbers satisfy:\,\,\,$n_{k}^{ij}+n_{ij}=n_{j}^{ki}+n_{ki}$.  As pointed out in Ref.~\cite{Levinsen201407}, these coefficients may be related to Wigner's $d$-matrix~\cite{Wigner1959} which assists in their simplification;\,\,\,for example \footnote{The corresponding equation in Ref.~\cite{Levinsen201407} quoted the angle as $2\pi/3$, however with the coordinates defined cyclically, it should have read $4\pi/3$.  Nonetheless, the results of Ref.~\cite{Levinsen201407} are unchanged, as this factor of two only leads to a sign change for those Clebsch-Gordan coefficients which multiply the relative wave function $\varphi_n$ with $n$ odd (i.e., such terms do not contribute for short-range $s$-wave interactions).}:
\begin{align} \label{eq:TSU(3)F-09}
\langle 0,n_{1}^{23},n_{23}|0,n_{2}^{31},n_{31}\rangle= 
d_{\frac{n_{2}^{31}-n_{31}}2,\frac{n_{1}^{23}-n_{23}}2}^{\frac{n_{1}^{23}+n_{23}}2}\left(4\pi/3\right).
\end{align}

Equation~\eqref{eq:TSU(3)F-07} implies that the spectrum may be obtained by finding the eigenvalues of the matrix on the right-hand side, i.e., this matrix depends on the energy, and the eigenvalues equal the inverse coupling constant.  The system is solved numerically for different values of $E$ by increasing the maximum value of the ``atom-atom'' quantum number in the sums, Eqs.~\eqref{eq:TSU(3)F-08A} and \eqref{eq:TSU(3)F-08B}, and of the ``atom-pair'' quantum number, until each of the eigenvalues $1/g$ converges.  Note that since the relevant ``atom-atom'' relative motion is always even for short-range $s$-wave interactions, the spectrum decouples into sectors of even and odd ``atom-pair'' harmonic oscillator quantum numbers.

The energy spectrum is displayed in Fig.~\ref{fig:TSU(3)F}, where we have omitted the bound states (which only exist for attractive interactions) to focus on the so-called upper branch of the spectrum.  We note that there are no avoided crossings between bound and upper-branch states.  We can see how the energies increase for increasing repulsion, until at infinite coupling the spectrum becomes highly degenerate.  At this point, the system is fermionized in the sense that the probability density is equivalent to that of non-interacting identical fermions~\cite{Girardeau196011}.  For instance, the ground-state energy of the upper branch for $g\rightarrow\infty$ is $E=0+1+2=3$, which is exactly the same as for 3 identical fermions.

To gain further insight into the spectrum, we also solve the Schr\"{o}dinger equation for other three-body systems, namely $3$ identical bosons and $2+1$ distinguishable fermions (i.e., 2 identical fermions and a distinguishable particle).  In both of these problems, the Hamiltonian is the same as for SU($3$) fermions.  However, for identical bosons, the wave function is symmetric under pair exchange, and consequently, we have $\eta_{12}=\eta_{23}=\eta_{31}\equiv \eta$.  Equation~\eqref{eq:TSU(3)F-07} then reduces to
\begin{align} \label{eq:TSU(3)F-11}
\frac{1}{g}\eta=A\eta+2B\eta\,.
\end{align} 
It is important to emphasize here that the additional complexity in the SU($3$) problem with distinguishable particles is because, upon permuting the atoms, the relative signs of $\{\phi_{\hat{P}\{\n\}}\}$ are undefined.  As a result, we obtain three coupled equations in terms of three independent $\eta$'s, rather than a single equation using one $\eta$.

For the problem of $2+1$ fermions, identical particles are unaffected by the contact interactions because the wave function is antisymmetric under their exchange.  If we say that atoms with coordinates $z_{1}$ and $z_{2}$ are spin-$\uparrow$, and the atom with coordinate $z_{3}$ is spin-$\downarrow$, then we have the relation, $\phi_{n_{1}n_{2}n_{3}}=-\phi_{n_{2}n_{1}n_{3}}$, which leads to the expression,
\begin{align} \label{eq:TSU(3)F-12}
\frac{1}{g}\eta=A\eta-B\eta\,.
\end{align}
The solutions afforded by Eqs.~\eqref{eq:TSU(3)F-11} and~\eqref{eq:TSU(3)F-12} are both contained in Fig.~\ref{fig:TSU(3)F}, and this can be explained using group theory, as we now discuss below.

As we have said, the Hamiltonians for different three-body systems are exactly the same.  The only difference is in the quantum statistics that restrict the permutation symmetries of the wave functions, which can be represented by Young diagrams.  Consider taking the outer product (denoted by $\otimes$) of the irreducible representations for $2$ identical fermions and another fermion, respectively denoted by the Young diagrams\,\,\,${\small\Yvcentermath1 \yng(1,1)}$\,\,\,and\,\,\,${\small\Yvcentermath1 \yng(1)}$\,.  The second graph can be added to the first ``in all possible ways''~\cite{Hamermesh1989}, and we find that the resulting representation is a direct sum of two irreducible representations of the permutation group S$_3$.  This means that the eigenstates for $2+1$ fermions include those for 3 identical fermions:
\begin{align} \label{eq:TSU(3)F-13A}
{\small\Yvcentermath1 \yng(1,1)\,\,\otimes\,\,\yng(1)\,\,=\,\,\yng(1,1,1)\,\,\oplus\,\,\yng(2,1)}\,\,.
\end{align}
Likewise, $2$ identical bosons\,\,\,${\small\Yvcentermath1 \yng(2)}$\,\,\,and an additional boson\,\,\,${\small\Yvcentermath1 \yng(1)}$\,\,\,contain the states for $3$ identical bosons:
\begin{align} \label{eq:TSU(3)F-13B}
{\small\Yvcentermath1 \yng(2)\,\,\otimes\,\,\yng(1)\,\,=\,\,\yng(2,1)\,\,\oplus\,\,\yng(3)}\,\,.
\end{align}
Now, in the same way, we see that the wave function for three distinguishable SU($3$) fermions resolves into the states for $3$ identical fermions, $2+1$ fermions, $2+1$ bosons, and $3$ identical bosons~\cite{Hamermesh1989, Harshman201601A, Decamp201605}:
\begin{align} \label{eq:TSU(3)F-14}
{\small\Yvcentermath1 \yng(1)\,\,\otimes\,\,\yng(1)\,\,\otimes\,\,\yng(1)\,\,=\,\,\yng(1,1,1)\,\,\oplus\,\,2\,\,\yng(2,1)\,\,\oplus\,\,\yng(3)}\,\,.
\end{align}
The line above describes the decomposition of the Hilbert space with respect to the permutation symmetry.  In other words, the symmetries of the states in the SU(3) spectrum correspond to the irreducible representations of S$_{3}$, and their degeneracies are given by the dimensions of those representations~\cite{Hamermesh1989}.  Notice that the number of states which transform according to a particular Young tableau is also equal to the degeneracy.  All of these features are clearly illustrated in Fig.~\ref{fig:TSU(3)F}.  This is a very similar idea to how Young diagrams have been used previously to decompose the few-body two-component Fermi gas into different subsystems \cite{Daily2012}.

To summarize, in this section, we have used the three-body case to explain the method by which we solve the Schr\"{o}dinger equation for small SU($N$) Fermi gases, and we have shown how the solutions project onto both bosonic and fermionic sectors.  Subsequently, we extend this approach to the four-body problem.

\section{Four-Body Problem}
\label{sec:Four-Body_Problem}

We move on to consider four distinguishable SU($4$) fermions that are harmonically confined in one dimension.  This problem can be solved by employing the same techniques that were introduced in the previous section.  Here, we note that various methods for the numerical treatment of two-component four-fermion systems already exist in the literature:\,\,\,see, for example, Refs.~\cite{Gharashi201307, Sowinski201309, Volosniev201411,Pecak201705}.

\begin{figure*}[ht]
\begin{center}
\includegraphics[scale=0.4]{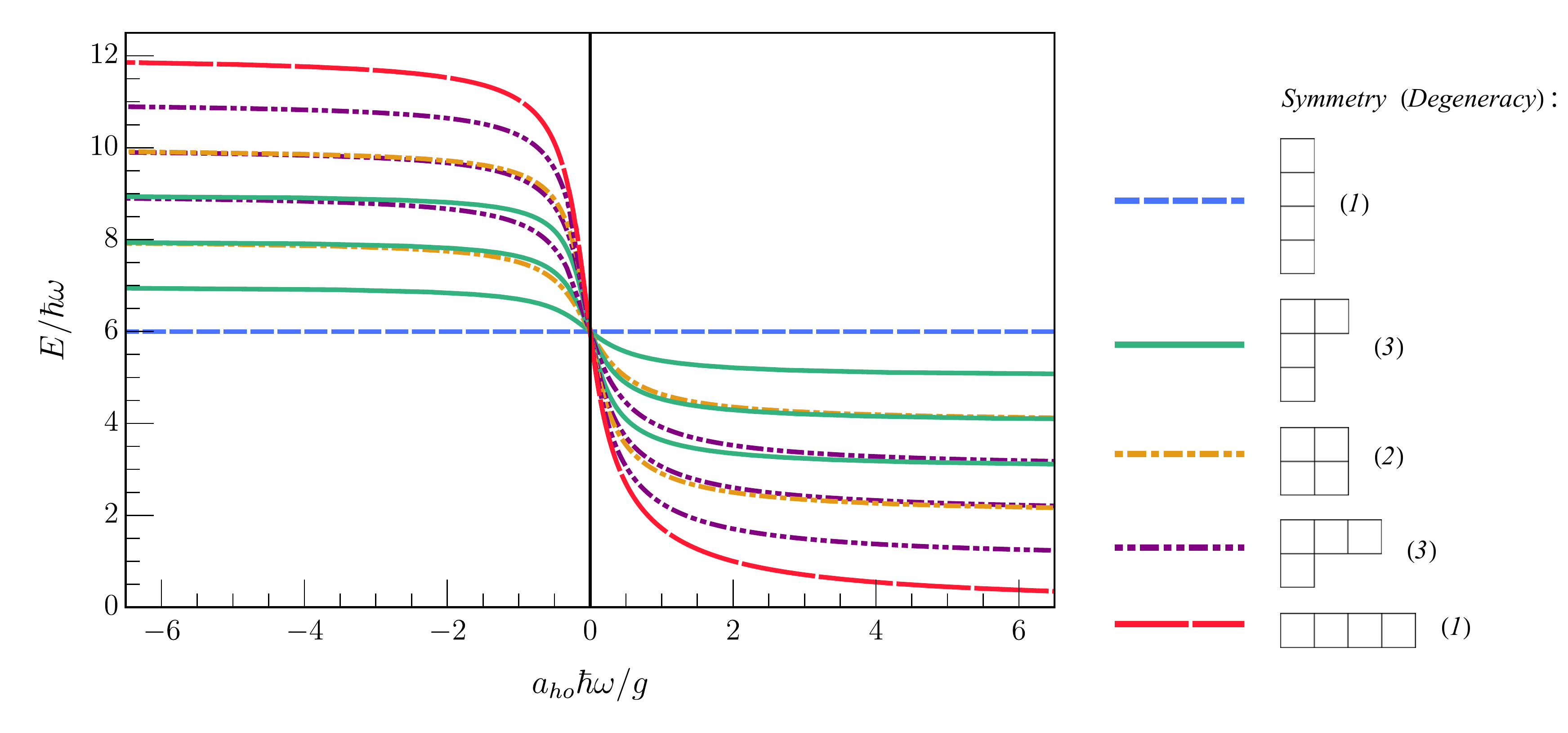}
\caption{The energy spectrum of the ground-state manifold for four distinguishable SU($4$) fermions interacting in a one-dimensional harmonic trap.  The energies $E$ are obtained as solutions to Eqs.~\eqref{eq:FSU(4)F-05}-\eqref{eq:FSU(4)F-new}, and are shown according to their dependence on the inverse interaction strength $1/g$.  This system projects onto one- and two-component Bose and Fermi gases, corresponding to the irreducible representations of S$_{4}$ (see main text).}
\label{fig:FSU(4)F}
\end{center}
\end{figure*}

We start by writing down the general four-body wave function,
\begin{align} \label{eq:FSU(4)F-02}
|\psi^{(4)}\rangle=\sum_{\n}\phi_{\n}|\n\rangle\,,
\end{align}
where $\n\equiv\{n_{1},\,n_{2},\,n_{3},\,n_{4}\}$.  Once again, we project the Schr\"{o}dinger equation onto the non-interacting eigenstate $|\n\rangle$, and then change to the relative and centre-of-mass coordinates of the particles via the transformation:
\begin{align} \label{eq:FSU(4)F-03}
\left(\begin{array}{c}
z_{ij} \\ [4pt]
z_{kl} \\ [4pt]
z_{kl}^{ij} \\ [4pt]
z_{\textit{cm}} \\ [4pt]
\end{array}\right)
=
\left(\begin{array}{ccccc}
1 & -1 & 0 & 0 \\ [4pt]
0 & 0 & 1 & -1 \\ [4pt]
\frac{1}{2} & \frac{1}{2} & -\frac{1}{2} & -\frac{1}{2} \\ [4pt]
\frac{1}{4} & \frac{1}{4} & \frac{1}{4} & \frac{1}{4} \\ [4pt]
\end{array}\right)
\left(\begin{array}{c}
z_{i} \\ [4pt]
z_{j} \\ [4pt]
z_{k} \\ [4pt]
z_{l} \\ [4pt]
\end{array}\right),
\end{align}
where $\{i,\,j,\,k,\,l\}=\{1,\,2,\,3,\,4\}$ and permutations with $i<j,\,k$, and $k<l$.  The coordinate $z_{ij}$ ($z_{kl}$) describes the relative motion of atoms $i$ and $j$ ($k$ and $l$), $z_{kl}^{ij}$ is the relative motion between their two centers of mass, and $z_{\textit{cm}}$ is the center-of-mass motion of all four atoms (see Fig.~\ref{fig:sketch}).  This leads us to define six independent wave functions by subjecting Eq.~\eqref{eq:FSU(4)F-02} to the boundary condition that two particles, $i$ and $j$ (or $k$ and $l$), are on top of each other:
\begin{align} \label{eq:FSU(4)F-04}
\eta_{ij} & \equiv\eta_{(ij)}^{n_{kl}^{ij},n_{kl}}=g\sum_{n_{ij},\,\n}\varphi_{n_{ij}}\langle0,n_{kl}^{ij},n_{kl},n_{ij}|\n\rangle\phi_{\n}\,, \nn \\
\eta_{kl} & \equiv\eta_{(kl)}^{n_{kl}^{ij},n_{ij}}=g\sum_{n_{kl},\,\n}\varphi_{n_{kl}}\langle0,n_{kl}^{ij},n_{kl},n_{ij}|\n\rangle\phi_{\n}\,.
\end{align}
Here, $\{i,\,j,\,k,\,l\}$ take the same values as above, and again we let $n_{\textit{cm}}=0$.  By carrying out manipulations similar to those in the three-body problem, we obtain six coupled equations in terms of these wave functions:
\begin{align}
&\frac{1}{g}
\left(\begin{array}{c}
\eta_{12} \\
\eta_{34} \\
\eta_{13} \\
\eta_{24} \\
\eta_{14} \\
\eta_{23} \\
\end{array}\right)
=\nn \\
&
\begin{pmatrix}
A & B & C & C^{(1)} & C^{(3)} & C^{(4)} \\
B & A & C^{(1)} & C & C^{(1)} & C \\
C & C^{(1)} & A & B & C^{(2)} & C^{(4)} \\
C^{(1)} & C & B & A & \left[C^{(4)}\right]^T & C \\
\left[C^{(3)}\right]^T & C^{(1)} & C^{(2)} & C^{(4)} & A & B\\
\left[C^{(4)}\right]^T & C & \left[C^{(4)}\right]^T & C & B & A\\
\end{pmatrix}
\left(\begin{array}{c}
\eta_{12} \\
\eta_{34} \\
\eta_{13} \\
\eta_{24} \\
\eta_{14} \\
\eta_{23} \\
\end{array}\right),\label{eq:FSU(4)F-05}
\end{align}
where $\{A,\,B,\,C\}$ are real, symmetric tensors given by
\begin{align} \label{eq:FSU(4)F-06}
A_{n,l;n',l'}=\sum_{p}\frac{\varphi_{p}^{2}}{E-p-n-l}\delta_{n,n'}\delta_{l,l'}\,,
\end{align}
\begin{align} 
B_{n,l;n',l'}=\frac{\varphi_l\varphi_{l'}}{E-n-l-l'}\delta_{n,n'}
\end{align}
and
\begin{align} \label{eq:FSU(4)F-07}
C_{n,l;n',l'} = \sum_{p,\,p'} \frac{\varphi_{p} \varphi_{p'}}{E-p-n-l} \widetilde{T}_{n',l',p'}^{n,l,p}\,. 
\end{align}
Here, the four-body Clebsch-Gordan coefficient $\widetilde{T}_{n',l',p'}^{n,l,p}$ is defined via the transformation,
\begin{align} \label{eq:FSU(4)F-CGCex}
|0,n_{34}^{12},n_{34},n_{12}\rangle = 
\sum_{n_{24}^{13},\,n_{24},\,n_{13}}
\widetilde{T}_{n_{24}^{13},n_{24},n_{13} }^{n_{34}^{12},n_{34},n_{12}}
|0,n_{24}^{13},n_{24},n_{13}\rangle\,.
\end{align}
The remaining tensors are simply related to $C$ as follows:
\begin{align} \label{eq:FSU(4)F-new}
C_{n,l;n',l'}^{(1)} & = (-1)^{n+l'} C_{n,l;n',l'}\, , \nn \\ 
C_{n,l;n',l'}^{(2)} & = (-1)^{l+l'} C_{n,l;n',l'}\, , \nn \\
C_{n,l;n',l'}^{(3)} & = (-1)^{l} C_{n,l;n',l'} \, ,\nn \\
C_{n,l;n',l'}^{(4)} & = (-1)^{n'} C_{n,l;n',l'} \, .
\end{align}
In this SU(4) problem, the ``pair-pair'' $n_{kl}^{ij}$ and ``atom-atom'' $n_{kl}$ quantum numbers give rise to the matrix structure in Eqs.~\eqref{eq:FSU(4)F-06}-\eqref{eq:FSU(4)F-07}.  That is to say, just like in the SU(3) case, the numerics now depend on two fewer parameters than what we started with:\,\,\,$\{n_{kl},\,n_{kl}^{ij}\}$ instead of $\{n_{1},\,n_{2},\,n_{3},\,n_{4}\}$.  The four-body equations decouple into even and odd ``atom-atom plus pair-pair'' $n_{kl}+n_{kl}^{ij}$ motion, which allows us to speed up their numerical evaluation.  This furthermore implies that $C^{(1)}$ is a symmetric tensor in each odd or even $n_{kl}+n_{kl}^{ij}$ sector.

From group theory considerations~\cite{Hamermesh1989, Harshman201601B, Decamp201605}, we know that four distinguishable SU($4$) fermions project onto $4$, $3+1$ and $2+2$ systems with either bosonic or fermionic character.  These symmetries are associated with the irreducible representations of the permutation group S$_{4}$, as shown below:
\begin{align} \label{eq:FSU(4)F-08}
&{\small\Yvcentermath1 \yng(1)\,\,\otimes\,\,\yng(1)\,\,\otimes\,\,\yng(1)\,\,\otimes\,\,\yng(1)=} \\ \nonumber
&{\small\Yvcentermath1 \yng(1,1,1,1)\,\,\oplus\,\,3\,\,\yng(2,1,1)\,\,\oplus\,\,2\,\,\yng(2,2)\,\,\oplus\,\,3\,\,\yng(3,1)\,\,\oplus\,\,\yng(4)}\,\,.
\end{align}
Again, both the degeneracy of a given state and the number of states transforming according to a particular Young tableau, are given by the dimensionality of the representation.

In Fig.~\ref{fig:FSU(4)F}, we show the ground-state manifold of the upper-branch spectrum for the four-body problem, obtained by numerically solving Eqs.~\eqref{eq:FSU(4)F-05}-\eqref{eq:FSU(4)F-new}.  In the limit of weak repulsive interactions, the ground-state energy is determined by the representation under which the corresponding wave function transforms.  The horizontal line depicts the fully antisymmetric state, which is not affected by the interparticle interactions.  As $g$ increases towards infinite repulsion, the energies of the other interacting states increase until they all become degenerate at $g\rightarrow\infty$, where the energy $E=0+1+2+3=6$ is the same as for 4 identical fermions.  As $g$ decreases from infinite attraction, the degeneracy is again lifted and the states continue to smoothly increase in energy.  Notice how the curvatures of the energies are not symmetric around the $1/g\rightarrow0$ axis, but are slightly steeper on the attractive side compared to the repulsive side.  The colouring scheme and legend help to illustrate the group theoretical aspect of these results.

We are now done with demonstrating the technique by which we solve the Schr\"{o}dinger equation for few-body SU($N$) Fermi gases.  In the next section, we examine the limit of near infinite interactions by mapping the system Hamiltonian to a spin chain.

\section{Strong-Coupling Limit}
\label{sec:Strong-Coupling_Limit}

\bgroup
\def\arraystretch{1.5}
\begin{table*}[t]
\begin{center}
\begin{tabular}{|>{\hspace{0.1pc}}c<{\hspace{0.1pc}}|>{\hspace{0.1pc}}c<{\hspace{0.1pc}}|>{\hspace{0.1pc}}c<{\hspace{0.1pc}}|>{\hspace{0.1pc}}c<{\hspace{0.1pc}}|>{\hspace{0.1pc}}c<{\hspace{0.1pc}}|}
\hline
\multirow{2}{*}{Subsystem} & \multirow{2}{*}{Degeneracy} & \multicolumn{2}{c|}{Contact, $\mathcal{C}_{n}/\mathcal{C}_{1}$} & Maximum Overlap, \\
\cline{3-4}
& & Ansatz, $|\widetilde{\psi}\rangle$ & Exact, $|\psi\rangle$ & $|\langle\psi|\widetilde{\psi}\rangle|_{\mathrm{max}}$ \\
\hline
\multicolumn{1}{|l|}{\hspace{0.39cm}{\tiny\Yvcentermath1$\yng(1,1,1,1)$}} & $1$ & $0$ & $0$ & $1$ \rule{0pt}{4.0ex} \\
\multicolumn{1}{|l|}{\hspace{0.39cm}{\tiny\Yvcentermath1$\yng(2,1,1)$}} & $3$ & $1$ & $1$ & $0.999993$ %\rule{0pt}{3.0ex}
\\
\multicolumn{1}{|l|}{\hspace{0.39cm}{\tiny\Yvcentermath1$\yng(2,2)$}} & $2$ & $5-\sqrt{7}$ & $0.99548\times(5-\sqrt{7})$ & $0.999983$ \\
\multicolumn{1}{|l|}{\hspace{0.39cm}{\tiny\Yvcentermath1$\yng(2,1,1)$}} & $3$ & $3$ & $1.00632\times3$ & $1$ \rule{0pt}{3.0ex} \\
\multicolumn{1}{|l|}{\hspace{0.39cm}{\tiny\Yvcentermath1$\yng(3,1)$}} & $3$ & $4$ & $1.00474\times4$ & $0.999993$ \\
\multicolumn{1}{|l|}{\hspace{0.39cm}{\tiny\Yvcentermath1$\yng(2,1,1)$}} & $3$ & $6$ & $0.99696\times6$ & $0.999993$ %\rule{0pt}{3.0ex}
\\
\multicolumn{1}{|l|}{\hspace{0.39cm}{\tiny\Yvcentermath1$\yng(3,1)$}} & $3$ & $7$ & $0.99739\times7$ & $1$ \\
\multicolumn{1}{|l|}{\hspace{0.39cm}{\tiny\Yvcentermath1$\yng(2,2)$}} & $2$ & $5+\sqrt{7}$ & $1.00148\times(5+\sqrt{7})$ & $0.999983$ \\
\multicolumn{1}{|l|}{\hspace{0.39cm}{\tiny\Yvcentermath1$\yng(3,1)$}} & $3$ & $9$ & $1.00008\times9$ & $0.999993$ \\
\multicolumn{1}{|l|}{\hspace{0.39cm}{\tiny\Yvcentermath1$\yng(4)$}} & $1$ & $10$ & $1.00007\times10$ & $1$ \\
\hline
\end{tabular}
\caption{\,\,\,Extending the ansatz of Ref.~\cite{Levinsen201507} to four-component fermions.  We consider the 24 states that comprise the ground-state manifold for four distinguishable SU($4$) fermions in the limit of strong repulsive interactions ($1/g\to0^+$).  The contact $\mathcal{C}_{n}$, Eq.~\eqref{eq:SCL-02}, for the $n^{th}$ eigenstate is given by the $n^{th}$ eigenvalue of the exchange Hamiltonian $\hat{H}'$, Eq.~\eqref{eq:SCL-01}.  For clarity, we consider only distinct contacts and we normalize these relative to $\mathcal{C}_{1}$, defined as the contact of the first state as we go down in energy from the fermionized state (which has $\mathcal{C}_{0}=0$).  Because the trapping potential is symmetric, there are two unique values for the nearest-neighbour exchange coefficients, $J_{1}=J_{3}$ and $J_{2}$.  The analytic coefficients have been taken from Ref.~\cite{Loft201605}, while the ansatz coefficients ($J_{1}=1.755$ and $J_{2}=2.340$) can be calculated using Eq.~\eqref{eq:SCL-03} with $\kappa_{4}=0.59588$ \cite{Massignan201512}.  Also shown are the maximum overlaps $|\langle\psi|\widetilde{\psi}\rangle|_{\mathrm{max}}$ between the ansatz $|\widetilde{\psi}\rangle$ and exact $|\psi\rangle$ wave functions associated with each energy slope, which may be determined from Eq.~\eqref{eq:SCL-04}.}
\label{table:Table_I}
\end{center}
\end{table*}
\egroup

In the strong-coupling limit, the problem of SU($N$) fermions in one dimension can be mapped to a ``spin-exchange'' model.  The idea~\cite{Deuretzbacher201407} is to think of the system in terms of an effective one-dimensional lattice, where the first atom is at the site $i=1$, the second atom is at the site $i=2$, and so on.  In the limit where the coupling constant diverges ($1/g=0$), the atoms are impenetrable and, for a given ordering of particles, the ground-state wave function is proportional to that of ${\cal N}$ identical fermions (i.e., a Slater determinant of the lowest ${\cal N}$ single-particle eigenfunctions).  Such fermionization of two distinguishable fermions has recently been observed in experiment~\cite{Zurn201202}.  As we perturb away from that limit, adjacent atoms can swap places, and thus we introduce a nearest-neighbour exchange interaction to the picture for large but finite $g$.  This type of mapping has been successfully applied to two-component Bose and Fermi gases~\cite{Matveev200403, Matveev200810, Deuretzbacher201407, Volosniev201411, Volosniev201502, Levinsen201507, Massignan201512}, to Fermi gases with both $s$- and $p$-wave coupling~\cite{Yang2016}, and to multi-component systems~\cite{Deuretzbacher201407,Yang2015,Pan201702}.

For SU($N$) fermions that are described by Eq.~\eqref{eq:INTRO-01}, the exchange Hamiltonian has the form~\cite{Deuretzbacher201407}:
\begin{align} \label{eq:SCL-01}
\hat{H} & \simeq E_{\infty} - \frac{1}{g} \hat{H}' \nn \\
& = E_{\infty} + \frac{1}{g} \sum_{i\,=\,1}^{\mathcal{N}-1} J_{i} \left( \hat{P}_{i,\,i+1} - 1 \right) .
\end{align}
Above, the first term $E_{\infty}$ is the unperturbed Hamiltonian, which gives the ``fermionized'' energy (equivalent to the energy of ${\cal N}$ identical fermions), and the second term contains the perturbation $\hat{H}'$ to first order in $1/g$, which gives the exchange interaction energy.  The summation index $i$ denotes the effective lattice site, while $\mathcal{N}$ is the total number of particles, and the operator $\hat{P}_{i,\,i+1}$ permutes pairs of adjacent atoms, with an energy cost that is determined by the nearest-neighbour exchange constant $J_{i}$.  The values of these constants depend on the nature of the confining potential.  (Note that only permutations of distinguishable particles give non-zero contributions to the Hamiltonian.)  Using perturbation theory and the Hellmann-Feynman theorem~\cite{Volosniev201411}, it can be shown that the (negative) energy slope of the $n^{th}$ eigenstate in the ground-state manifold is given by the corresponding eigenvalue of the perturbation.  Expressly,
\begin{align} \label{eq:SCL-02}
- \frac{dE_n}{d(g^{-1})} \Bigr|_{g\rightarrow\infty}
&= -\left<\frac{\partial\hat{H}}{\partial(g^{-1})} 
\right>_{\psi_n}\Bigr|_{g\rightarrow\infty} 
\nn \\ 
&= \frac{\langle \psi_{n} | \hat{H}' | \psi_{n} \rangle}{\langle \psi_{n} | \psi_{n} \rangle}
\equiv \mathcal{C}_{n} \, ,
\end{align}
where $\mathcal{C}_{n}$ is the one-dimensional contact density of the state $|\psi_{n}\rangle$.

Before solving the Hamiltonian $\hat{H}'$, we shall first discuss its symmetry properties.  Without loss of generality, we restrict our attention to the case of $N$ \emph{distinguishable} SU($N$) fermions, since its spectrum contains all the eigenstates of other $N$-particle systems.  Immediately, we can find a trivial eigenstate corresponding to the fermionic wave function,
\begin{align}
|\Psi_f\rangle = \sum_{\hat{P}} \hat{P} |\alpha_1,\,\alpha_2,\,\ldots,\,\alpha_N\rangle \, ,
\end{align}
where $\alpha_i$ with $i=1,\,2,\,\ldots,\,N$ is the spin index, and $\hat{P}$ represents a permutation of these indices.  We note that, while $|\Psi_f\rangle$ appears symmetric, the required antisymmetry is in the real-space part of the wave function. It is clear that $\hat{H}'|\Psi_f\rangle =0$, which means the energy is independent of the interaction strength $g$.  Based on this fermionic wave function, it is possible to construct another eigenstate $|\Psi_b\rangle$ corresponding to the bosonic state,
\begin{align}
|\Psi_b\rangle = \sum_{\hat{P}} \delta_{\hat{P}} \hat{P} |\alpha_1,\,\alpha_2,\,\ldots,\,\alpha_N\rangle \, ,
\label{eq:sym}
\end{align}
where $\delta_{\hat{P}}$ is the sign of the permutation $\hat{P}$.  Its corresponding eigenvalue is then given by $E_b=-2\sum_iJ_i$.  This is an example of how we can transform from one eigenstate with a certain permutation symmetry to another eigenstate with a different symmetry.

In fact, as discussed in Ref.~\cite{Sutherland197511}, there exists a mapping between any two conjugate irreducible representations of S$_N$.  Namely, if $|\Psi_R\rangle$ is an eigenstate in an irreducible representation $R$, then we can use it to construct another eigenstate $|\Psi_{\bar{R}}\rangle$ in the conjugate representation $\bar{R}$:
\begin{align}
& \ket{\Psi_{R}} = \sum_{\hat{P}} \chi_{\hat{P}} \hat{P} \ket{\alpha_1,\,\alpha_2,\,\ldots,\,\alpha_N} \nn \\
& \rightarrow \,\,\,\,\, \ket{\Psi_{\bar{R}}} = \sum_{\hat{P}} \delta_{\hat{P}} \chi_{\hat{P}} \hat{P} \ket{\alpha_1,\,\alpha_2,\,\ldots,\,\alpha_N} \, ,
\label{eq:MG}
\end{align}
where $\chi_{\hat{P}}$ is a coefficient that depends on the specific ordering of the spins along the effective lattice.  It is straightforward to check that $|\Psi_{R}\rangle$ and $|\Psi_{\bar{R}}\rangle$ are two states belonging to conjugate representations $R$ and $\bar{R}$.  Furthermore, their eigenenergies satisfy $E_R + E_{\bar{R}} = - 2 \sum_i J_i$, which implies that the spectrum of $\hat{H}'$ is symmetric around $- \sum_i J_i$.

Now, we progress to investigating the eigenstates of the Hamiltonian, Eq.~\eqref{eq:SCL-01}.  In the SU(2) case, this Hamiltonian can be mapped to an XXX Heisenberg spin chain~\cite{Deuretzbacher201407}.  Within this framework, an ansatz was recently introduced~\cite{Levinsen201507} which provides highly accurate analytical wave functions for a single impurity ($N_\down=1$) in a Fermi sea of $N_\up=\mathcal{N}-1$ majority atoms.  Using the ansatz, the exchange coefficients for a harmonic confinement were found to take the approximate analytical form,
\begin{align} \label{eq:SCL-03}
J_{i} = i (\mathcal{N} - i) \kappa_{\mathcal{N}} \, ,
\end{align}
where $i$ varies from $1$ to $\mathcal{N}-1$, and $\kappa_{\mathcal{N}}$ is an $\mathcal{N}$-dependent constant.  In Ref.~\cite{Levinsen201507}, it was reported that for $N_{\up}\leq8$, the overlap between the ansatz and exact wave functions in the ground-state manifold is larger than $0.9997$ for all states.  We have calculated the overlaps for $N_{\up}\leq29$ using the numerically exact coefficients which were subsequently published in Ref.~\cite{Loft201605}, and we find that all overlaps between exact and ansatz wave functions in the ground-state manifold exceed $0.9947$.  In Ref.~\cite{Massignan201512}, the ansatz was applied to spin-balanced two-component Bose gases, and again it was found that the resulting approximate wave functions were nearly identical to the numerically exact solutions.  Because the ansatz describes these problems extremely well, it is of interest to see whether it can be extended to SU($N$) Fermi gases.  To do this, we diagonalize the Hamiltonian of Eq.~\eqref{eq:SCL-01} using both the ansatz~\cite{Levinsen201507} and the numerically exact coefficients~\cite{Loft201605} to obtain a comparison.  We utilize the same constants $J_{i}$ for the $N$-component systems as for the two-component case, since they do not depend on the particles' spin, but only on their position in the effective lattice.

\begin{figure}[t]
\begin{center}
\includegraphics[scale=0.37]{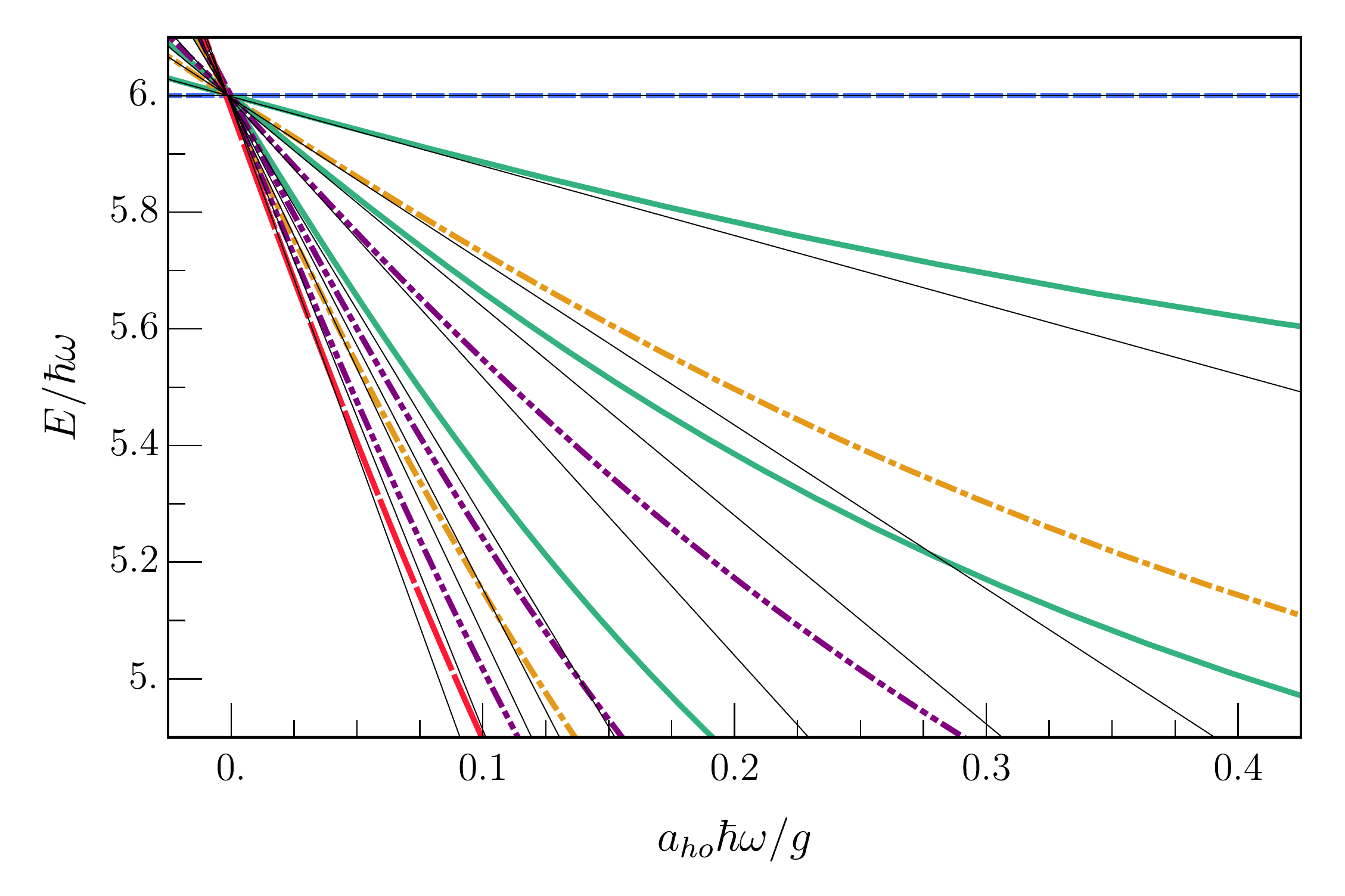}
\caption{A close-up view of the \mbox{$1/g\rightarrow0^+$} region of the ground-state energy spectrum for four distinguishable SU($4$) fermions with repulsive interactions.  The differently coloured/patterned curves are the exact energies also shown in Fig.~\ref{fig:FSU(4)F}, while overlaid as thin black lines are the energy slopes predicted by the ansatz~\cite{Levinsen201507} within the exchange model, Eqs.~\eqref{eq:SCL-01} and~\eqref{eq:SCL-03} (with $\kappa_{4}=0.59588$~\cite{Massignan201512}).}
\label{fig:SCL-01}
\end{center}
\end{figure}

While the two- and three-body problems are exactly described by the ansatz coefficients in Eq.~\eqref{eq:SCL-03}, the four-body problem is non-trivial.  Shown above in Table~\ref{table:Table_I} are our findings for four distinguishable SU($4$) fermions, which was the system considered in Sec.~\ref{sec:Four-Body_Problem}.  Remarkably, we find that the approximate energy slopes (third column) are almost identical to the exact slopes (fourth column) for all 10 non-degenerate energy levels.  This indicates that the ansatz~\cite{Levinsen201507} near exactly reproduces the energy levels for the entire manifold.  To highlight this result, we have superimposed the slopes given by the ansatz onto the actual energy spectrum in Fig.~\ref{fig:SCL-01}.

As well as the energy slopes, we also compare the exact and approximate wave functions (in what follows, denoted $|\psi\rangle$ and $|\widetilde{\psi}\rangle$) by calculating their maximum overlap $|\langle\psi|\widetilde{\psi}\rangle|_{\mathrm{max}}$.  For a given degenerate eigenvalue, the wave function is any normalized, linear superposition of the $d$ degenerate eigenvectors.  In other words, $|\psi\rangle$ and $|\widetilde{\psi}\rangle$ are each found in a $d$-dimensional subspace of a 24-dimensional state space.  The key here is that, without loss of generality, we can find $|\langle\psi|\widetilde{\psi}\rangle|_{\mathrm{max}}$ just by determining the maximum overlap between $|\psi\rangle$ and any one of the eigenvectors for $|\widetilde{\psi}\rangle$.  Furthermore, this particular $|\psi\rangle$ is orthogonal to all the remaining eigenvectors for $|\widetilde{\psi}\rangle$.  Since each degenerate subspace is an irreducible representation of the permutation group, we can find these orthogonal eigenvectors by applying a proper linear combination of the permutation onto $|\psi\rangle$ and $|\widetilde{\psi}\rangle$.  It follows that the maximum overlap is, in fact, very easy to calculate and given by:
\begin{align} \label{eq:SCL-04}
|\langle\psi|\widetilde{\psi}\rangle|_{\mathrm{max}}
=
(\,|\,\mathrm{det}[\,
{\cal O}
\,]\,|\,)^{1/d}\,,
\end{align}
where ${\cal O}$ is the overlap matrix ${\cal O}_{ij}=\langle\psi_{i}|\widetilde{\psi}_{j}\rangle$ of eigenstates within the degenerate subspaces, with $1\leq i,\,j\leq d$.
For two SU($2$) fermions and three SU($3$) fermions, the degree of agreement between the ansatz and exact wave functions is always $100$\,\%.  The values for the four-body problem are included in Table~\ref{table:Table_I}, where we see that $|\langle\psi|\widetilde{\psi}\rangle|_{\mathrm{max}}$ remains either exactly equal to or very close to one.  The equality of some of the overlaps arises either from the transformation in Eq.~\eqref{eq:MG}, or from parity considerations~\cite{Levinsen201507}.

\begin{figure}[t]
\begin{center}
\includegraphics[scale=0.38]{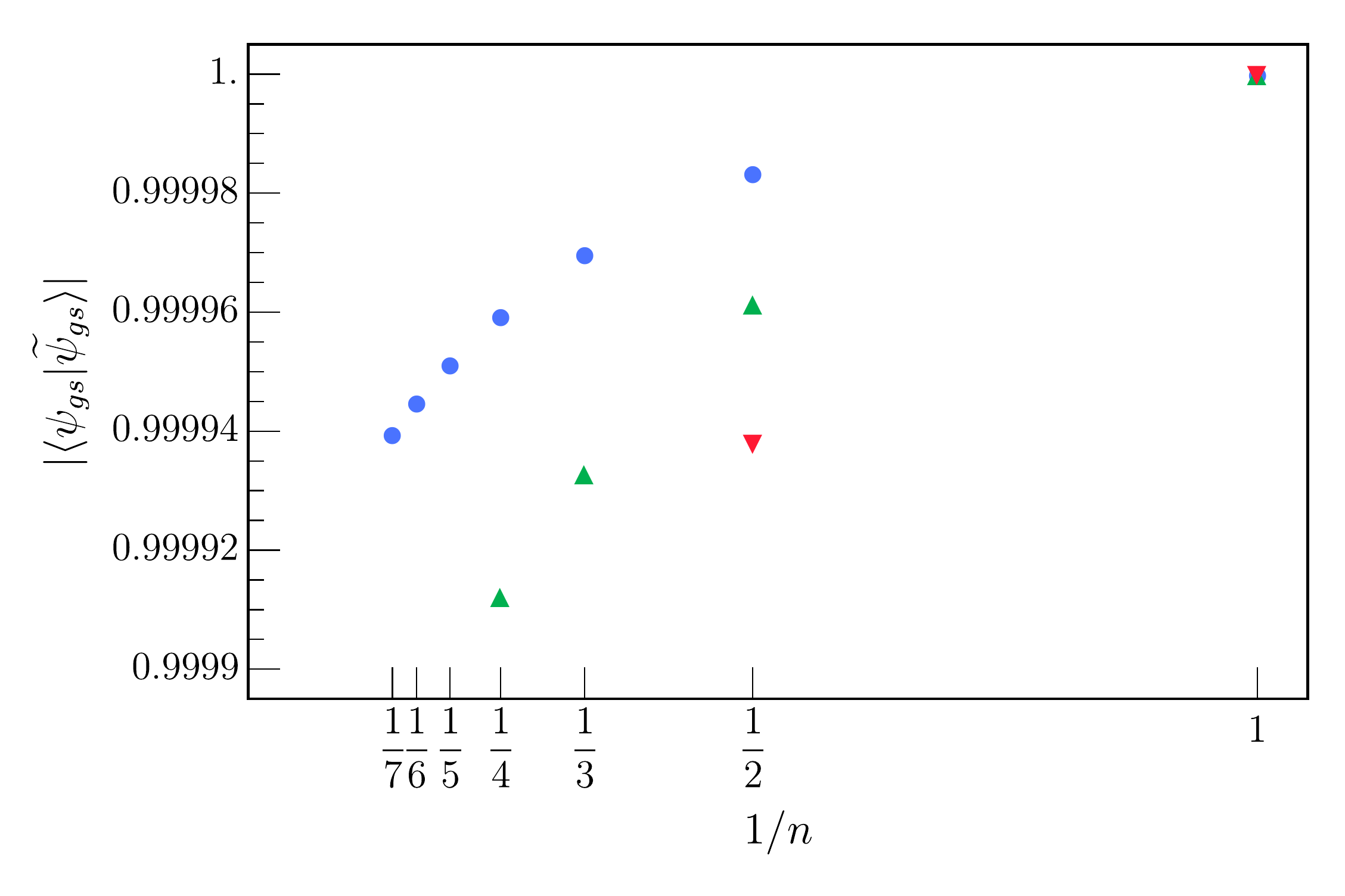}
\caption{The overlap between the ansatz $|\widetilde{\psi}_{gs}\rangle$ and exact $|\psi_{gs}\rangle$ ground-state wave functions for balanced SU($2$) [blue circle,\,\,\mycircle{blue}\,], SU($3$) [green up-triangle,\,\,\myuptriangle{green}\,] and SU($4$) [red down-triangle,\,\,\mydowntriangle{red}\,] Fermi gases, with an increasing number of particles $n$ in each component.  The three systems share a concurrent point at $(1,1)$.}
\label{fig:SCL-02}
\end{center}
\end{figure}

As a further test of the ansatz, we compare the approximate $|\widetilde{\psi}_{gs}\rangle$ and exact $|\psi_{gs}\rangle$ ground-state wave functions for balanced SU($N=2,\,3,\,4$) Fermi gases, where there is an increasing number of atoms $n$ in each spin state.  The overlaps are presented in Fig.~\ref{fig:SCL-02}.  When there is just one particle of each spin, the ground state is fully symmetric (see Eq.~\eqref{eq:sym}) and the ansatz is exact:\,\,\,$|\langle\psi_{gs}|\widetilde{\psi}_{gs}\rangle|=1$.  For more particles --- up to seven when $N=2$, and four when $N=3$ --- the overlaps remain above $99.99$\,\%.  These results, combined with the other findings of this section, show that the strong-coupling ansatz of Ref.~\cite{Levinsen201507} can be successfully extended from two- to $N$-component fermions.  This yields great advantage when it comes to studying larger-$N$ systems, because the ansatz coefficients in Eq.~\eqref{eq:SCL-03} are so simple to calculate.

\section{Rapid Convergence of the Ground-State Energy}
\label{sec:Rapid_Convergence_of_the_Ground-State_Energies}

\begin{figure*}[ht]
\begin{center}
\includegraphics[scale=0.38]{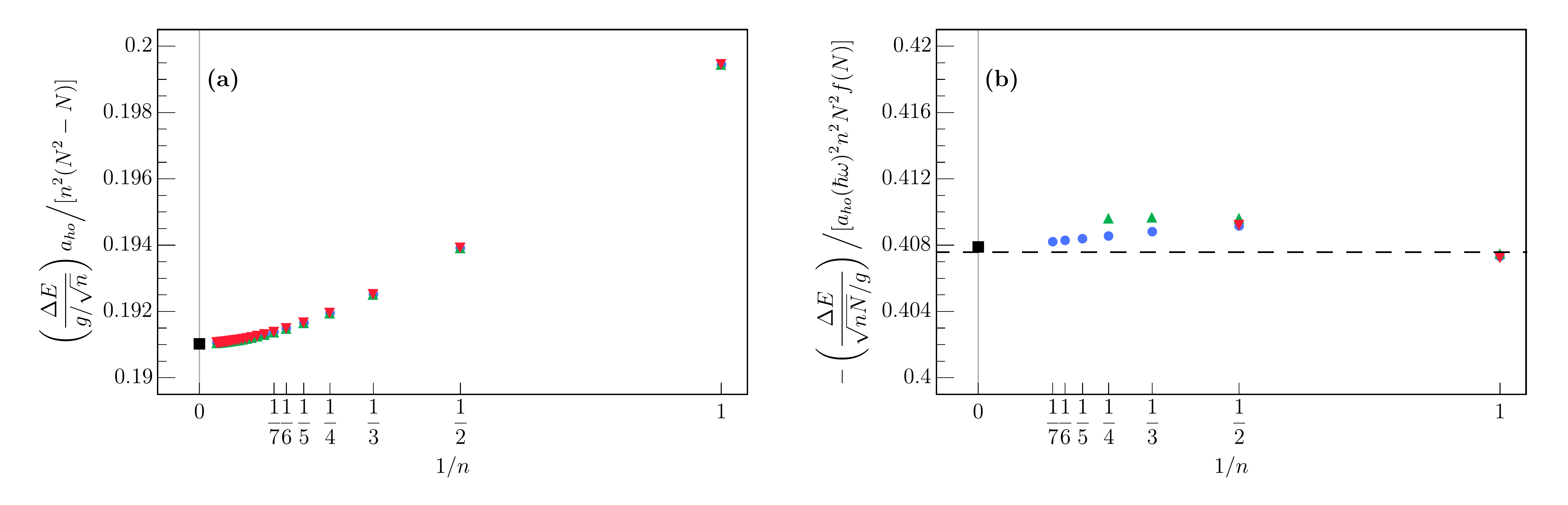}
\caption{The rescaled ground-state energy slopes $\Delta E$ for SU($2$) [blue circle,\,\,\mycircle{blue}\,], SU($3$) [green up-triangle,\,\,\myuptriangle{green}\,] and SU($4$) [red down-triangle,\,\,\mydowntriangle{red}\,] Fermi gases, with $n$ particles in each component.  Plots \textbf{(a)} and \textbf{(b)} are the weak- and strong-coupling slopes, respectively, obtained using first-order perturbation theory.  In panel \textbf{(a)}, we show the many-body limit as a black square, and data points are plotted for up to $30$ particles per spin state.  In panel \textbf{(b)}, we normalize the slopes by the function $f(N)=1-0.04/N-1.45/N^{2}+0.60/N^{3}$, and the limit for $N\to\infty$ with $n=1$ is indicated by the dashed line.  The $n\to\infty$ limit of the SU(2) system is shown as a black square.}
\label{fig:RC}
\end{center}
\end{figure*}

Systems of many particles are, in general, very difficult to describe theoretically owing to their many degrees of freedom.  One approach is to study such systems from the few-body limit.  For one-dimensional SU(2) Fermi gases, the energy of the ground state has been shown to converge to the many-body result for very small particle numbers in both theory~\cite{Gharashi201501, Grining201512} and experiment~\cite{Wenz201310}.  Similarly, both the Bertsch parameter and the contact of the unitary three-dimensional Fermi gas have been found to converge very quickly to their many-body limits~\cite{Levinsen2017}.  This provides strong motivation for few-fermion studies, and the question arises whether the rapid convergence towards the many-body limit is replicated in SU($N$) Fermi systems with $N$ larger than 2.  Reference~\cite{Matveeva201606} has already reported that over a large range of interaction strengths, the contact and interaction energy (measured in appropriate units) are quite insensitive to the number of particles in each component $n$, and only depend on the number of spin flavors $N$.  Here, we complement this result by looking in detail at the weak- and strong-coupling limits of the repulsive SU($N$) Fermi gas.

As done in the previous section, we consider balanced systems of SU($N=$ $2,\,3,\,4$) fermions with $n$ atoms of each spin.  Using perturbation theory to first order, we calculate the normalized ground-state energy slopes in the limits of weak and strong interactions, and examine how they change as $n$ is increased.

In the weak repulsive limit ($g\rightarrow0^+$), each of the $N$ components can be treated as an ideal Fermi gas, i.e., the wave function $|\psi_{gs}\rangle$ is found simply by taking the outer product of $N$ Slater determinants.  When we perturb the system by introducing non-zero interactions, the first-order correction to the energy is equal to the expectation value of the perturbation --- $\hat{H}_{\mathrm{int}}$ in Eq.~\eqref{eq:INTRO-01} --- with respect to the unperturbed wave function:\,\,\,$\Delta E=\langle\psi_{gs}|g\sum_{i\,<\,j}\delta(z_{i}-z_{j})|\psi_{gs}\rangle$. With $n$ particles in each of the $N$ spin states, this expression takes the form:
\begin{align} \label{eq:RC-01}
\Delta E=\frac{g}\pi\sum_{r\,=\,0}^{n-1}\sum_{s\,=\,0}^{n-1}\frac{N(N-1)}{2^{r+s+1}r!s!}\int_{-\infty}^{\infty}dx\,e^{-2x^{2}}H_{r}^{2}(x)H_{s}^{2}(x)\,,
\end{align}
which is precisely $N(N-1)/2$ times the corresponding two-component energy shift.  In the limit $n\to\infty$, this reduces to the expected weak-coupling result obtained using the local density approximation,\,\,\,$\Delta E=4\sqrt{2}\,g\,N(N-1)n^{3/2}/(3\pi^{2})$~\cite{Astrakharchik200512}.  Our results are shown in Fig.~\ref{fig:RC}\textbf{(a)}, where we see that there is a fast convergence to the many-body value of $\Delta E$, with a finite-$n$ correction of at most 5\,\%.  Moreover, in this limit, we know the dependence on $N$ exactly.

In the opposite limit of strong repulsion, the energy shift is instead\,\,\,$\Delta E=-\frac1g\langle\psi_{gs}|\hat H'|\psi_{gs}\rangle$, leading to the results shown in Fig.~\ref{fig:RC}\textbf{(b)}.  Here we note that, for increasing $N$, the ground-state energy shift quickly approaches that of an ideal Bose gas with ${\cal N}=n\times N$ particles.  Within the local density approximation applied to the Lieb-Liniger solution, this has the thermodynamic limit,\,\,\,$\Delta E=-\frac{128\sqrt{2}}{45\pi^2}(nN)^{5/2}/g$~\cite{Astrakharchik200512} (shown as a dashed line).  In other words, like for weak-coupling, the ground-state energy shift appears to be predictable with good accuracy from the limit of $N$ identical bosons (i.e., the ground-state energy for $N$ spin components with $n=1$), and the number of particles $n$ in each component appears to matter little.  Furthermore, we note that in the thermodynamic limit of $n\to\infty$, the SU(2) system has\,\,\,$\Delta E=-\frac{128\sqrt{2}\ln2}{45\pi^2}(2n)^{5/2}/g$~\cite{Astrakharchik200512}, which we see is extremely close to the Lieb-Liniger limit.  Since it is reasonable to believe that the ground-state wave function becomes increasingly symmetric with increasing number of spin components, we conjecture that the thermodynamic limit for any SU($N$) system with $N\geq3$ will lie in the very small interval between these two limits.  In order to reduce finite-$N$ effects, in Fig.~\ref{fig:RC}\textbf{(b)}, we have normalized by the function $f(N)$, which is obtained by fitting to the $n=1$ ground-state energies for $N\leq30$.

The fact that both perturbative limits converge quickly with increasing $n$, suggests that the full ground-state spectra are converging to the many-body energies at a fast rate.  This complements the conclusions drawn in Ref.~\cite{Matveeva201606}, where the ground-state contact for $N$-component fermions was found to be characterized by the same rapid convergence.  These results are also consistent with those of Ref.~\cite{Yang2011}, where it was shown that the ground-state energy per particle of repulsive SU($N$) gases in uniform one-dimensional space reduces to that of spinless bosons in the $N\to\infty$ limit.  Together, our works imply that the energetics of balanced, many-body SU($N=2,\,3,\,4$) Fermi gases in the ground state are well described by the corresponding few-body systems of $N=2,\,3,\,4$ \emph{distinguishable} particles.

\section{Conclusions and Outlook}
\label{sec:Conclusions_and_Outlook}

Motivated by recent experiments on $^{173}$Yb atoms, we have undertaken a theoretical study of SU($N$) Fermi gases under a one-dimensional harmonic confinement.  Initially, we considered the case of a single particle in each of the $N$ spin components.  For these systems, we presented a method for solving the Schr\"{o}dinger equation and obtaining the full numerically exact energy spectra.  The primary advantage of our approach is that we are able to remove two harmonic oscillator indices from the equations, which greatly simplifies the numerics.  We solved the problems of three and four distinguishable SU($N$) fermions, and we believe that our technique can be adapted to the five- and six-body problems, which are yet to be solved in the literature.  In particular, the five-body SU(5) system is the smallest that will feature solutions not present in two-component Fermi or Bose gases.

We then specialized to studying the ground-state manifold of the upper-branch spectra near the limit of infinite interactions.  There, the Hamiltonian for SU($N$) Fermi gases can be mapped to a spin-exchange model~\cite{Deuretzbacher201407}.  Within this setting, we applied an ansatz~\cite{Levinsen201507}, which was originally devised for the fermion impurity problem, and showed that it can be extended to $N$-component fermions.  As in the two-component case~\cite{Levinsen201507,Massignan201512}, the ansatz is exceedingly accurate in this application.  Finally, we considered SU($N=2,\,3,\,4$) Fermi gases with an equal number of particles, $n$, in each spin state.  Using first-order perturbation theory, we found that the ground-state energy slopes, in both the weak- and strong-coupling limits, converge rapidly to the many-body results as $n$ is increased from one.  This suggests, in agreement with other reports~\cite{Wenz201310, Gharashi201501, Grining201512, Matveeva201606}, that the energetics of multi-component many-body Fermi systems may be accurately estimated from the few-body regime.

Our results open up the possibility of using few-body approaches to gain insight into the behavior of SU($N$) systems.  For instance, one could address spin and pairing correlations in one-dimensional lattices~\cite{Capponi201604};\,\,\,indeed, antiferromagnetic correlations have already been observed in a one-dimensional microtrap containing a few spin-$\up$ and $\down$ fermions~\cite{Murmann201511, Deuretzbacher2017}.  One can also consider large-spin fermionic systems in higher dimensions, where exotic spin correlations and magnetic phase transitions are expected to be observed~\cite{Marston198906, Read199009, Harada200303, Assaad200502, Kawashima200701, Xu200804, Hermele200909, Cazalilla200910, Gorshkov201002, Cazalilla201411, Cichy201605}.  Finally, there is the prospect of using few-body energy spectra to determine the thermodynamic properties at finite temperature.

\acknowledgments

We gratefully acknowledge discussions with Pietro Massignan and Xi-Wen Guan at the outset of this work.  We also thank Pietro Massignan for useful comments on the manuscript.  This work was performed in part at the Aspen Center for Physics, which is supported by the National Science Foundation Grant No. PHY-1607611.  J.L. is supported through the Australian Research Council Future Fellowship FT160100244.  Z.Y.S., M.M.P. and J.L. also acknowledge financial support from the Australian Research Council via Discovery Project No.~DP160102739.

\onecolumngrid

\appendix
\section*{Appendix:\,\,\,\,\,\,\,Four-Body Clebsch-Gordan Coefficients}

We give the derivation of the four-body Clebsch-Gordan coefficients appearing in Eq.~\eqref{eq:FSU(4)F-05}.  As an example, we look at\,\,\,$\widetilde{T}_{\,n_{24}^{13},n_{24},n_{13}}^{\,n_{34}^{12},n_{34},n_{12}} \equiv \langle 0,n_{34}^{12},n_{34},n_{12}|0,n_{24}^{13},n_{24},n_{13}\rangle$, which is defined in Eq.~\eqref{eq:FSU(4)F-CGCex}.

We start with the non-interacting part of the Hamiltonian, Eq. \eqref{eq:INTRO-01}, which may be re-written as
\begin{align} \label{eq:A-01}
\hat{H}_{0}=\sum_{i\,=\,1}^{4}\left[-\frac{\hbar^{2}}{2m}\frac{\partial^{2}}{\partial z_{i}^{2}}+\frac{m\omega^{2}}{2}z_{i}^{2}\right]=-\frac{\hbar^{2}}{2}\,\partial_{\mathbf{z}}^{T}\,\textbf{M}^{-1}\,\partial_{\mathbf{z}}+\frac{\omega^{2}}{2}\,\mathbf{z}^{T}\,\textbf{M}\,\mathbf{z}\,,
\end{align}
where $\partial_{\mathbf{z}}^{T}=\{\partial_{z_{1}},\,\partial_{z_{2}},\,\partial_{z_{3}},\,\partial_{z_{4}}\}$ and $\mathbf{z}^{T}=\{z_{1},\,z_{2},\,z_{3},\,z_{4}\}$ are vectors, and $\textbf{M}=m\textbf{I}$ is the mass matrix (with $\textbf{I}$ being the $4\times4$ identity matrix).  Clearly, Eq.~\eqref{eq:A-01} forms a set of four decoupled harmonic oscillators.  The eigenfunctions of the harmonic oscillator have energies $E_{n}=(n+\frac{1}{2})\hbar\omega$, and are given by
\begin{align} \label{eq:A-02}
\psi_{n}^{(i)}(z)=\frac{1}{\sqrt{2^{n}n!}}\left(\frac{m_{i}\omega}{\pi\hbar}\right)^{\frac{1}{4}}\mathrm{exp}\left(-\frac{m_{i}\omega }{2\hbar}\,z^{2}\right)H_{n}\left(\sqrt{\frac{m_{i}\omega}{\hbar}}\,z\right),
\end{align}
where $H_{n}(x)$ are the Hermite polynomials.  As done in Sec.~\ref{sec:Four-Body_Problem}, we define the following transformation which takes us from the single-particle basis to the relative motion basis:
\begin{align} \label{eq:A-03}
\left(\begin{array}{c}
z_{ij} \\
z_{kl} \\
z_{kl}^{ij} \\
Z \\
\end{array}\right)
=
\left(\begin{array}{cccc}
1 & -1 & 0 & 0 \\
0 & 0 & 1 & -1 \\
1/2 & 1/2 & -1/2 & -1/2 \\
1/4 & 1/4 & 1/4 & 1/4 \\
\end{array}\right)
\left(\begin{array}{c}
z_{i} \\
z_{j} \\
z_{k} \\
z_{l} \\
\end{array}\right)
=
\textbf{U}_{ij}\,\textbf{z}\,,
\end{align}
where $\{i,\,j,\,k,\,l\}=\{1,\,2,\,3,\,4\}$ and permutations with $i<j,\,k$, and $k<l$.  The transformed Hamiltonian has the same form as Eq.~\eqref{eq:A-01}, except that $\textbf{M}$ is replaced by the transformed mass matrix:
\begin{align} \label{eq:A-04}
\textbf{M}_{ij}
=
[(\textbf{U}_{ij})^{-1}]^{\mathrm{T}}\cdot\textbf{M}\cdot(\textbf{U}_{ij})^{-1}
=
m
\left(\begin{array}{cccc}
\frac{1}{2} & 0 & 0 & 0 \\
0 & \frac{1}{2} & 0 & 0 \\
0 & 0 & 1 & 0 \\
0 & 0 & 0 & 4 \\
\end{array}\right)
=
\left(\begin{array}{cccc}
m_{1,1} & 0 & 0 & 0 \\
0 & m_{1,1} & 0 & 0 \\
0 & 0 & m_{2,2} & 0 \\
0 & 0 & 0 & m_{4} \\
\end{array}\right).
\end{align}
Here, $m_{1,1}$ is the relative mass of two particles, $m_{2,2}$ is the relative mass of two pairs of particles, and $m_{4}$ is the total mass of all four atoms.  The harmonic oscillator eigenfunctions that correspond to this transformation are denoted $\psi_{n_{ij}}^{(1,1)}(z_{ij})$, $\psi_{n_{kl}}^{(1,1)}(z_{kl})$, $\psi_{n_{kl}^{ij}}^{(2,2)}(z_{kl}^{ij})$ and $\psi_{N}^{(4)}(Z)$, according to Eq.~\eqref{eq:A-02}.

Now, we recall that the Hermite polynomials are given by an exponential generating function,
\begin{align} \label{eq:A-05}
\mathrm{exp}\,(2xt-t^{2})=\sum_{n\,=\,0}^{\infty}H_{n}(x)\,\frac{t^{n}}{n!}\,,
\end{align}
and set up the following equality:
\begin{multline} \label{eq:A-06}
\mathrm{exp}\left(2\,\frac{z_{12}}{\alpha_{1,1}}\,t_{1}-t_{1}^{2}\right)\,\mathrm{exp}\left(2\,\frac{z_{34}}{\alpha_{1,1}}\,t_{2}-t_{2}^{2}\right)\,\mathrm{exp}\left(2\,\frac{z_{34}^{12}}{\alpha_{2,2}}\,t_{3}-t_{3}^{2}\right)= \\ \mathrm{exp}\left(2\,\frac{z_{13}}{\alpha_{1,1}}\,t_{1}'-(t_{1}')^{2}\right)\,\mathrm{exp}\left(2\,\frac{z_{24}}{\alpha_{1,1}}\,t_{2}'-(t_{2}')^{2}\right)\,\mathrm{exp}\left(2\,\frac{z_{24}^{13}}{\alpha_{2,2}}\,t_{3}'-(t_{3}')^{2}\right),
\end{multline}
where $\alpha_{i}=\sqrt{\hbar/m_{i}\omega}$ is the characteristic length for $\psi_{n}^{(i)}(z)$.  The vectors $\textbf{t}^{T}=\{t_{1},\,t_{2},\,t_{3}\}$ and $(\textbf{t'})^{T}=\{t_{1}',\,t_{2}',\,t_{3}'\}$ are related by a unitary transformation, which allows us to write down
\begin{align} \label{eq:A-07}
(\textbf{U}_{12}\,\textbf{z})^{T}\sqrt{\textbf{M}_{12}}\,\textbf{t}=(\textbf{U}_{13}\,\textbf{z})^{T}\sqrt{\textbf{M}_{13}}\,\textbf{t'}
\end{align}
from Eq.~\eqref{eq:A-06}.  Rearranging for $\textbf{t'}$, we find:
\begin{align} \label{eq:A-08}
\left(\begin{array}{c}
t_{1}' \\
t_{2}' \\
t_{3}' \\
\end{array}\right)
=
\left(\begin{array}{cccc}
\frac{1}{2} & -\frac{1}{2} & \frac{1}{\sqrt{2}} \\
-\frac{1}{2} & \frac{1}{2} & \frac{1}{\sqrt{2}} \\
\frac{1}{\sqrt{2}} & \frac{1}{\sqrt{2}} & 0 \\
\end{array}\right)
\left(\begin{array}{c}
t_{1} \\
t_{2} \\
t_{3} \\
\end{array}\right)
\equiv
\left(\begin{array}{cccc}
a & b & c \\
d & e & f \\
g & h & i \\
\end{array}\right)
\left(\begin{array}{c}
t_{1} \\
t_{2} \\
t_{3} \\
\end{array}\right).
\end{align}
Inserting Eq.~\eqref{eq:A-05} into Eq.~\eqref{eq:A-06}, and using Eq.~\eqref{eq:A-08}, gives
\begin{multline} \label{eq:A-09}
\sum_{n_{12},\,n_{34},\,n_{34}^{12}}H_{n_{12}}\left(\frac{z_{12}}{\alpha_{1,1}}\right)H_{n_{34}}\left(\frac{z_{34}}{\alpha_{1,1}}\right)H_{n_{34}^{12}}\left(\frac{z_{34}^{12}}{\alpha_{2,2}}\right)\frac{t_{1}^{n_{12}}\,t_{2}^{n_{34}}\,t_{3}^{n_{34}^{12}}}{n_{12}!\,n_{34}!\,n_{34}^{12}!}
= \\
\sum_{n_{13},\,n_{24},\,n_{24}^{13}}H_{n_{13}}\left(\frac{z_{13}}{\alpha_{1,1}}\right)H_{n_{24}}\left(\frac{z_{24}}{\alpha_{1,1}}\right)H_{n_{24}^{13}}\left(\frac{z_{24}^{13}}{\alpha_{2,2}}\right)\sum_{j_{1}+j_{2}+j_{3}\,=\,n_{13}}\,\sum_{k_{1}+k_{2}+k_{3}\,=\,n_{24}}\,\sum_{l\,=\,0}^{n_{24}^{13}}\binom{n_{13}}{j_{1},\,j_{2},\,j_{3}}\binom{n_{24}}{k_{1},\,k_{2},\,k_{3}}\binom{n_{24}^{13}}{l} \\
\times\, a^{j_{1}}b^{j_{2}}c^{j_{3}}d^{k_{1}}e^{k_{2}}f^{k_{3}}g^{l}h^{n_{24}^{13}-l}\,\frac{t_{1}^{j_{1}+k_{1}+l}\,t_{2}^{j_{2}+k_{2}-l+n_{24}^{13}}\,t_{3}^{j_{3}+k_{3}}}{n_{13}!\,n_{24}!\,n_{24}^{13}!}\,,
\end{multline}
with $0\leq j_{i}\leq n_{13}$ and $0\leq k_{i}\leq n_{24}$, where we have performed multinomial expansions on both $(t_{1}')^{n_{13}}$ and $(t_{2}')^{n_{24}}$, and a binomial expansion on $(t_{3}')^{n_{24}^{13}}$.  By equating like powers of $t_{i}$ and building the harmonic oscillator eigenfunctions around the Hermite polynomials, we obtain
\begin{multline} \label{eq:A-10}
\psi_{n_{12}}^{(1,1)}(z_{12})\,\psi_{n_{34}}^{(1,1)}(z_{34})\,\psi_{n_{34}^{12}}^{(2,2)}(z_{34}^{12})
=
\sum_{n_{13},\,n_{24},\,n_{24}^{13}}\sqrt{\frac{n_{12}!\,n_{34}!\,n_{34}^{12}!}{n_{13}!\,n_{24}!\,n_{24}^{13}!}}\sum_{j_{1}+j_{2}+j_{3}\,=\,n_{13}}\,\sum_{k_{1}+k_{2}+k_{3}\,=\,n_{24}}\,\sum_{l\,=\,0}^{n_{24}^{13}} \\
\binom{n_{13}}{j_{1},\,j_{2},\,j_{3}}\binom{n_{24}}{k_{1},\,k_{2},\,k_{3}}\binom{n_{24}^{13}}{l}\,a^{j_{1}}b^{j_{2}}c^{j_{3}}d^{k_{1}}e^{k_{2}}f^{k_{3}}g^{l}h^{n_{24}^{13}-l}\,\delta_{n_{12}\,,\,j_{1}+k_{1}+l}\,\delta_{n_{34}\,,\,j_{2}+k_{2}-l+n_{24}^{13}}\,\delta_{n_{34}^{12}\,,\,j_{3}+k_{3}} \\
\times\,
\psi_{n_{13}}^{(1,1)}(z_{13})\,\psi_{n_{24}}^{(1,1)}(z_{24})\,\psi_{n_{24}^{13}}^{(2,2)}(z_{24}^{13})\,.
\end{multline}
Here, we have also used the energy conservation laws,\,\,\,$n_{12}+n_{34}+n_{34}^{12}=n_{13}+n_{24}+n_{24}^{13}$\,\,\,and\,\,\,$(\omega/2)[m_{1,1}(z_{12})^{2}+m_{1,1}(z_{34})^{2}+m_{2,2}(z_{12}^{34})^{2}]=(\omega/2)[m_{1,1}(z_{13})^{2}+m_{1,1}(z_{24})^{2}+m_{2,2}(z_{13}^{24})^{2}]$.  The Clebsch-Gordan coefficient allows us to pass from the product of wave functions associated with the right-hand side of the coefficient to the product of those associated with the left hand side, i.e.,
\begin{align} \label{eq:A-11}
\psi_{n_{12}}^{(1,1)}(z_{12})\,\psi_{n_{34}}^{(1,1)}(z_{34})\,\psi_{n_{34}^{12}}^{(2,2)}(z_{34}^{12})
=
\sum_{n_{13},\,n_{24},\,n_{24}^{13}}\widetilde{T}_{\,n_{24}^{13},n_{24},n_{13}}^{\,n_{34}^{12},n_{34},n_{12}}\,\psi_{n_{13}}^{(1,1)}(z_{13})\,\psi_{n_{24}}^{(1,1)}(z_{24})\,\psi_{n_{24}^{13}}^{(2,2)}(z_{24}^{13})\,.
\end{align}
Thus, substituting in the values for $a$ through $h$, we can identify the four-body Clebsch-Gordan coefficient as
\begin{multline} \label{eq:A-12}
\widetilde{T}_{\,n_{24}^{13},n_{24},n_{13}}^{\,n_{34}^{12},n_{34},n_{12}}
=
2^{-n_{24}^{13}/2}\sqrt{\frac{n_{12}!\,n_{34}!\,n_{34}^{12}!}{n_{13}!\,n_{24}!\,n_{24}^{13}!}}\sum_{j_{1}+j_{2}+j_{3}\,=\,n_{13}}\,\sum_{k_{1}+k_{2}+k_{3}\,=\,n_{24}}\,\sum_{l\,=\,0}^{n_{24}^{13}}\binom{n_{13}}{j_{1},\,j_{2},\,j_{3}}\binom{n_{24}}{k_{1},\,k_{2},\,k_{3}}\binom{n_{24}^{13}}{l} \\
\times\,
(-1)^{j_{2}+k_{1}}\,2^{-(j_{1}+j_{2}+k_{1}+k_{2})-(j_{3}+k_{3})/2}\,\delta_{n_{12}\,,\,j_{1}+k_{1}+l}\,\delta_{n_{34}\,,\,j_{2}+k_{2}-l+n_{24}^{13}}\,\delta_{n_{34}^{12}\,,\,j_{3}+k_{3}}\,.
\end{multline}
This concludes our example.

The other coefficients in the four-body problem are equal to Eq.~\eqref{eq:A-12} up to a minus sign.  For instance, for the different coefficient, $\widetilde{T}_{\,n_{23}^{14},n_{23},n_{14}}^{\,n_{34}^{12},n_{34},n_{12}}\equiv\langle 0,n_{34}^{12},n_{34},n_{12}|0,n_{23}^{14},n_{23},n_{14}\rangle$, the transformation between $\textbf{t}$ and $\textbf{t'}$ is
\begin{align} \label{eq:A-13}
\left(\begin{array}{c}
t_{1}' \\
t_{2}' \\
t_{3}' \\
\end{array}\right)
=
\left(\begin{array}{cccc}
\frac{1}{2} & \frac{1}{2} & \frac{1}{\sqrt{2}} \\
-\frac{1}{2} & -\frac{1}{2} & \frac{1}{\sqrt{2}} \\
\frac{1}{\sqrt{2}} & -\frac{1}{\sqrt{2}} & 0 \\
\end{array}\right)
\left(\begin{array}{c}
t_{1} \\
t_{2} \\
t_{3} \\
\end{array}\right),
\end{align}
where $\{t_{1},\,t_{2},\,t_{3}\}$ are associated with $\{n_{12},\,n_{34},\,n_{34}^{12}\}$ and $\{t_{1}',\,t_{2}',\,t_{3}'\}$ are associated with $\{n_{14},\,n_{23},\,n_{23}^{14}\}$.  This means that we can write down
\begin{align} \label{eq:A-14}
\widetilde{T}_{\,n_{23}^{14},n_{23},n_{14}}^{\,n_{34}^{12},n_{34},n_{12}}
=
(-1)^{n_{34}}\,
\widetilde{T}_{\,n_{24}^{13},n_{24},n_{13}}^{\,n_{34}^{12},n_{34},n_{12}}\,.
\end{align}
The relationships shown in Eq.~\eqref{eq:FSU(4)F-new} of the main text may be deduced in a similar way.

\twocolumngrid

\bibliography{Bibliography}

\onecolumngrid

\end{document}